\documentclass[prd,twocolumn,eqsecnum,showpacs,amsfonts,amssymb]{revtex4}

\usepackage{graphicx}

\usepackage{bm}

\newcommand{\beq}{\begin{equation}}
\newcommand{\eeq}{\end{equation}}
\newcommand{\beqs}{\begin{eqnarray}}
\newcommand{\eeqs}{\end{eqnarray}}

\newcommand{\drawsquare}[2]{\hbox{%
\rule{#2pt}{#1pt}\hskip-#2pt
\rule{#1pt}{#2pt}\hskip-#1pt
\rule[#1pt]{#1pt}{#2pt}}\rule[#1pt]{#2pt}{#2pt}\hskip-#2pt
\rule{#2pt}{#1pt}}
\newcommand{\fund}{\raisebox{-.5pt}{\drawsquare{6.5}{0.4}}}
\newcommand{\sym}{\raisebox{-.5pt}{\drawsquare{6.5}{0.4}}\hskip-0.4pt%
        \raisebox{-.5pt}{\drawsquare{6.5}{0.4}}}
\newcommand{\asym}{\raisebox{-3.5pt}{\drawsquare{6.5}{0.4}}\hskip-6.9pt%
        \raisebox{3pt}{\drawsquare{6.5}{0.4}}}

\begin{document}

\title{Study of the Renormalization-Group Evolution of ${\cal N}=1$ 
Supersymmetric Gauge Theories Using Pad\'e Approximants}

\author{Gongjun Choi and Robert Shrock}

\affiliation{C.N. Yang Institute for Theoretical Physics, Stony Brook
  University, Stony Brook, NY 11794}

\begin{abstract}

We study asymptotically free SU($N_c$) gauge theories with ${\cal N}=1$
supersymmetry, including the purely gluonic theory and theories with $N_f$
copies of a pair of massless chiral superfields in the respective
representations $R$ and $\bar R$ of SU($N_c$).  The cases in which $R$ is the
fundamental representation and the symmetric and antisymmetric rank-2 tensor
representation are considered. We calculate Pad\'e approximants to the beta
functions for these theories in the $\overline{\rm DR}$ scheme up to four-loop
order for the gluonic theory and up to three-loop order for the theories with
matter superfields and compare results for IR zeros and poles with results from
the NSVZ beta function.  Our calculations provide a quantitative measure, for
these theories, of how well finite-order perturbative results calculated in one
scheme reproduce properties of a known beta function calculated in a different
scheme.

\end{abstract}

\pacs{11.15.-q,11.10.Hi,11.30.Pb}

\maketitle


\section{Introduction} 
\label{intro}

A fundamental issue in quantum field theories is the question of how
accurately, the beta function, calculated to a finite loop order, describes the
renormalization-group (RG) evolution of the theory when probed on different
Euclidean energy/momentum scales $\mu$.  There are two related aspects to this
question.  First, higher-order terms modify the value of the beta function
calculated to a given order, and this modification generically increases
with the size of the interaction coupling.  Second, although the one-loop and
two-loop terms in a beta function are independent of the scheme used for
regularization and renormalization, the three-loop and higher-loop terms depend
on this scheme.  Therefore, when studying a given property of a beta function
at higher-loop order, it is necessary to ascertain how significant this scheme
dependence is.  A particularly appealing context in which to investigate
properties of the beta function and their scheme dependence is a supersymmetric
gauge theory, because of the strong constraints that the supersymmetry places
on the properties of the theory.  

In this paper we shall use Pad\'e methods to study the beta function of
vectorial, asymptotically free, ${\cal N}=1$ supersymmetric SU($N_c$) gauge
theories (at zero temperature and chemical potential). We investigate both the
purely gluonic supersymmetric Yang-Mills (SYM) gauge theory and theories with
matter content consisting of $N_f$ copies of massless chiral superfields
$\Phi_i$ and $\tilde\Phi_i$, $i=1,...,N_f$, which transform according to the
respective representations $R$ and $\bar R$ of SU($N_c$).  We consider the
cases where $R$ is the fundamental representation and where $R$ is the
symmetric and antisymmetric rank-2 tensor representation of SU($N_c$).  We
denote the running gauge coupling of the theory by $g=g(\mu)$ and define
$\alpha(\mu) = g(\mu)^2/(4\pi)$. (The argument $\mu$ will often be suppressed
in the notation.)  The beta function is $\beta_g = dg/dt$, where $dt=d\ln\mu$,
or equivalently,
\beq
\beta_\alpha \equiv \frac{d\alpha}{dt} = \frac{g}{2\pi} \, \beta_g \ . 
\label{betadef}
\eeq
The beta function thus describes how the gauge coupling increases from the deep
ultraviolet (UV) at large $\mu$ to the infrared (IR) at small $\mu$.  The
asymptotic freedom property guarantees that in the deep UV, $\alpha(\mu) \ll
1$, so one can calculate the properties of the theory and its beta function
reliably using perturbative methods.  Depending on whether or not the theory
contains matter chiral superfields, and if so, what $R$ and $N_f$ are, this UV
to IR evolution may be governed by an IR zero of the beta function.  If the
theory has an IR zero and if it occurs at sufficiently small coupling, it may
be an exact IR fixed point of the renormalization group; alternatively, if it
occurs at sufficiently large coupling, spontaneous chiral symmetry breaking may
occur, giving dynamical masses to some particles, so that these are integrated
out in the low-energy effective field theory applicable below the scale of
condensate formation.  into the IR is governed by a different beta function.
In our analysis, we will make use of the closed-form calculation of
$\beta_\alpha$ by Novikov, Shifman, Vainshtein, and Zakharov (NSVZ) in
\cite{nsvz} (see also \cite{sv}), denoted $\beta_{\alpha,NSVZ}$, which is exact
within the scheme used for its calculation.  This beta function exhibits a pole
at a certain value of the coupling \cite{nsvz,sv,ks} (see
Eq. (\ref{apole_nsvz}) below).  Furthermore, we will make use of a number of
exact results that have been obtained using effective holomorphic action
methods concerning the infrared properties of this theory
\cite{seiberg,susyreview}.  The beta function of an ${\cal N}=1$ supersymmetric
gauge theory with general chiral superfield matter content has been calculated
up to three-loop order \cite{b12susy,b3susy}, and, for the pure gluonic SYM
theory, up to four-loop order \cite{b4sym} in the dimensional reduction scheme
with minimal subtraction \cite{drbar}, denoted $\overline{\rm DR}$.  Using
these results, we calculate Pad\'e approximants to the beta function of the
pure gluonic SYM theory up to four-loop order and to the beta functions of the
theories with chiral superfields in the fundamental and rank-2 tensor
representations up to three-loop order. The theories with sufficiently large
matter superfield content have a perturbative IR zero in the beta function.  In
previous work in \cite{bfs} with T. Ryttov and in \cite{bc}-\cite{lnn}, we have
calculated properties of the beta function, including an IR zero, from the two
and three-loop beta function. In the present work we extend these studies in
several ways. Using our calculation of Pad\'e approximants for the SYM theory
and for theories with various matter superfield content, we address and answer
several questions: (i) how the value of the IR zero in the Pad\'e approximants
compares with the IR zero in the $\overline{\rm DR}$ and NSVZ beta functions,
for cases where such an IR zero is present; (ii) whether these Pad\'e
approximants to the $\overline{\rm DR}$ beta function exhibit a robust
indication of a pole, as in the NSVZ beta function; (iii) if the answer to
question (ii) is affirmative, whether this pole occurs at a value of $\alpha$
near to the value in the NSVZ beta function and, moreover, closer to the origin
than an IR zero (if the latter is present) and hence dominates the UV to IR
evolution.  Our calculations and analysis provide a quantitative measure, for
these various supersymmetric theories, of how well finite-order perturbative
results calculated in the $\overline{\rm DR}$ scheme reproduce the properties
of the NSVZ beta function.  Some related work is in Refs.
\cite{jackjones}-\cite{ss}.

This paper is organized as follows.  In Sect. II we discuss the beta function
and exact results on the properties of the theory. In Sect. III we calculate
and analyze Pad\'e approximants for the pure gluonic supersymmetric Yang-Mills
theory. Sect. IV is devoted to the corresponding calculation and analysis of
Pad\'e approximants for the theory with chiral superfields in the fundamental
and conjugate fundamental representation.  In Sect. V we investigate the theory
with chiral superfields in the rank-2 tensor and conjugate tensor
representations.  Our conclusions are given in Sect. VI.


\section{Beta Function and Exact Results} 
\label{betafunction}


\subsection{Beta Function} 

In this section we review some basic results on the beta function and also some
exact results that we will use for our analysis.  The beta function
(\ref{betadef}) has the series expansion
\beq
\beta_\alpha = -2\alpha \sum_{\ell=1}^\infty b_\ell \, a^\ell  
         = -2\alpha \sum_{\ell=1}^\infty \bar b_\ell \, \alpha^\ell \ , 
\label{beta}
\eeq
where 
\beq
a \equiv \frac{g^2}{16\pi^2} = \frac{\alpha}{4\pi} \ , 
\label{a}
\eeq
$b_\ell$ is the $\ell$-loop coefficient, and 
$\bar b_\ell = b_\ell/(4\pi)^\ell$ the reduced $\ell$-loop coefficient. 
The first two coefficients in the expansion (\ref{beta}), which are 
scheme-independent, are \cite{b12susy,casimir}
\beq
b_1 = 3C_A - 2T_f N_f
\label{b1susy}
\eeq
and
\beq
b_2=6C_A^2-4(C_A+2C_f)T_fN_f \ . 
\label{b2susy}
\eeq
In the commonly used $\overline{\rm DR}$ scheme, 
the three-loop coefficient is \cite{b3susy}
\beqs
b_3 & = & 21C_A^3 + 4(-5C_A^2-13C_AC_f+4C_f^2)T_fN_f \cr\cr
    & + & 4(C_A+6C_f)(T_fN_f)^2 \ . 
\label{b3susy}
\eeqs
For pure ${\cal N}=1$ supersymmetric Yang-Mills theory (with no matter chiral
superfields, i.e., $N_f=0$), the four-loop coefficient, $b_4$, has also been
calculated \cite{b4sym} and will be used in our analysis of this SYM theory
below.

If $N_f=0$, then $b_1 > 0$; as $N_f$ increases from zero, $b_1$ decreases
monotonically and passes through zero, reversing sign, at the value
$N_f=N_{f,b1z}$, where \cite{nfintegral}
\beq
N_{f,b1z} = \frac{3C_A}{2T_f} 
\label{nfb1z}
\eeq
(and the subscript $b1z$ stands for ``$b_1$ zero'').  If $N_c$ and $R$ are such
that $N_{f,b1z}$ is an integer and if $N_f=N_{f,b1z}$, so that $b_1=0$, then
$b_2$ has the negative value $-12C_AC_f$.  Hence, the requirement of asymptotic
freedom, which means $\beta < 0$ near the origin, is true (given the minus sign
that we have extracted in Eq. (\ref{beta})) if and only if $b_1 >
0$. Therefore, we restrict to
\beq
N_f < N_{f,b1z} \ . 
\label{nfmax}
\eeq

Similarly, for $N_f=0$, $b_2 > 0$, and as $N_f$ increases from zero, $b_2$ 
decreases monotonically and passes through zero, reversing sign, at the value
$N_f=N_{f,b2z}$, where
\beq
N_{f,b2z} = \frac{3C_A^2}{2T_f(C_A+2C_f)} \ . 
\label{nfb2z}
\eeq
For an arbitrary fermion representation $R$, $N_{f,b2z} < N_{f,b1z}$, so there
is always an interval $I$ in $N_f$ where $b_1 > 0$ and $b_2 < 0$. This interval
is $N_{f,b2z} < N_f < N_{f,b1z}$, i.e., 
\beq
I: \quad \frac{3C_A^2}{2T_f(C_A+2C_f)} < N_f < \frac{3C_A}{2T_f} \ . 
\label{nfinterval}
\eeq
For $N_f \in I$, the two-loop ($2\ell$) beta function has an IR zero at
\beqs
\alpha_{IR,2\ell} & = & -\frac{\bar b_1}{\bar b_2} = -\frac{4\pi b_1}{b_2} 
\cr\cr
    & = & \frac{2\pi(3C_A-2T_fN_f)}{2(C_A+2C_f)T_fN_f-3C_A^2} \ . 
\label{alfir2loop}
\eeqs
Clearly, if $N_f$ is too close to $N_{f,b2z}$, then $b_2$ is sufficiently small
that $\alpha_{IR,2\ell}$ is too large for this perturbative two-loop result 
to be reliable.  As noted, the two-loop beta function encodes the maximal
scheme-independent perturbative information about the theory. 

Given that $N_f \in I$ and that $\alpha_{IR,2\ell}$ is sufficiently small for
the perturbative analysis of the beta function to be reasonable, a natural next
step in the analysis of the UV to IR evolution of the theory is to examine the
three-loop beta function.  The three-loop beta function has two zeros away from
the origin, given by the equation $b_1+b_2a+b_3a^2=0$ or equivalently, 
$\bar b_1 + \bar b_2 \alpha + \bar b_3 \alpha^2=0$.  The solutions are 
\beq
\alpha = \frac{1}{2\bar b_3}\Big [ -\bar b_2 \pm
\sqrt{\bar b_2^2-4\bar b_1\bar b_3} \ \Big ] \ .
\label{alphazerogeneral}
\eeq
The smaller one of these two solutions is the one that will be relevant for our
analysis, and we label it as $\alpha_{IR,3\ell}$. 


\subsection{NSVZ Beta Function} 

A closed-form expression for the beta function was derived by Novikov, Shifman,
Vainshtein, and Zakharov in \cite{nsvz} and discussed further in
\cite{sv}; this is
\beq
\beta_{\alpha,NSVZ} = \frac{d\alpha}{dt} = - \frac{\alpha^2}{2\pi} \bigg [
\frac{b_1-2T_fN_f\gamma_m}{1-2C_A a} \bigg ] \ , 
\label{beta_nsvz}
\eeq
where $\gamma_m$ is the anomalous dimension of the fermion bilinear 
bilinear product $\psi_i^T C \tilde \psi_i$, or equivalently, 
$\bar\psi_i \psi_i$, of component fermion fields in the quadratic superfield 
operator product $\Phi_i \tilde\Phi_i$ (no sum on the flavor index $i$). 
As noted above, this is an exact result within the scheme used in \cite{nsvz}.
This anomalous dimension has the series expansion 
\beq
\gamma_m = \sum_{\ell=1}^\infty c_\ell \, a^\ell
   = \sum_{\ell=1}^\infty \bar c_\ell \, \alpha^\ell \ ,
\label{gamma}
\eeq
where $\bar c_\ell = c_\ell/(4\pi)^\ell$ is the $\ell$-loop series coefficient.
Only the one-loop coefficient $c_1$ is scheme-independent, and is
\beq
c_1 = 4C_f \ .
\label{c1}
\eeq

Given our restriction to asymptotically free supersymmetric gauge theories, 
$\beta_{\alpha,NSVZ}$ has a UV zero at $\alpha=0$.  For the pure gluonic SYM
theory, $\beta_{\alpha,NSVZ}$ has no IR zero; for theories with nonzero 
matter superfield content, it may or may not have an IR zero, depending on this
content.  This will be discussed further below.  As is evident from 
Eq. (\ref{beta_nsvz}), $\beta_{\alpha,NSVZ}$ has a pole at 
\beq
a_{pole,NSVZ} = \frac{\alpha_{pole,NSVZ}}{4\pi} = \frac{1}{2C_A} \ . 
\label{apole_nsvz}
\eeq
An important property of this pole is that its position is independent of $R$ 
and $N_f$. 


\subsection{General Result on IR Phase Properties}

A number of exact results have been established for this asymptotically free
supersymmetric gauge theory \cite{nsvz,sv,seiberg} 
(see also \cite{susyreview,ryttov07}).  We recall one property that is
particularly relevant to our present work: if $N_f$ is in the interval 
$N_{f,cr} < N_f < N_{f,b1z}$, where 
\beq
N_{f,cr} = \frac{3C_A}{4T_f} = \frac{N_{b1z}}{2} \ ,
\label{nfcr}
\eeq
i.e., explicitly, the interval 
\beq
I_{N_f,NACP}: \quad \frac{3C_A}{4T_f} < N_f < \frac{3C_A}{2T_f} \ , 
\label{nf_nacp}
\eeq
then the theory flows from weak coupling in the UV to a superconformal IR fixed
point.  The resultant theory is in a (deconfined) non-Abelian Coulomb phase
(NACP) without any spontaneous chiral symmetry breaking.  In Eqs. (\ref{nfcr})
and (\ref{nf_nacp}), it is understood that, physically, $N_f$ must be an
integer \cite{nfintegral}, so the actual values of $N_f$ in the NACP are
understood to be the integers that satisfy the inequality (\ref{nf_nacp}).


\section{${\cal N}=1$ Supersymmetric Yang-Mills Theory}
\label{sym}

In this section we study the case $N_f=0$, i.e., supersymmetric Yang-Mills
theory. We use Pad\'e approximants to the $n$-loop beta function with $2 \le n
\le 4$ calculated in the $\overline{\rm DR}$ scheme to investigate how the
properties of this beta function compare with those of the NSVZ
beta function. This comparison elucidates the question of how sensitive
these properties are to the scheme used for the calculation. 

In this SYM theory the beta function depends on $a$ and $C_A$ via the product
\beq
x \equiv C_A a \ . 
\label{xdef}
\eeq
or equivalently, $\xi \equiv C_A \alpha$.  Consequently, it is natural to
re-express the beta function in terms of this product as the expansion
variable.  We thus define
\beq
\beta_x \equiv \frac{dx}{dt} = C_A \frac{da}{dt} = 
\frac{C_A}{4\pi} \, \beta_\alpha \ .
\label{betax}
\eeq
Since $b_\ell \propto C_A^\ell$, we define 
\beq
\hat b_\ell \equiv \frac{b_\ell}{C_A^\ell} \ . 
\label{bellhat}
\eeq
From Eqs. (\ref{b1susy})-(\ref{b3susy}), one has the two scheme-independent
coefficients
\beq
\hat b_1=3 \ , \quad \hat b_2=6
\label{b12hat}
\eeq
and, in the $\overline{\rm DR}$ scheme,
\beq
\hat b_3=21 \ .
\label{b3hat}
\eeq
For this SYM theory, the four-loop coefficient has also
been calculated in the $\overline{\rm DR}$ scheme \cite{b4sym}, and it is 
\beq
\hat b_4=102 \ . 
\label{b4hat}
\eeq
The beta function can be written as 
\beq
\beta_x = -2x^2 \, \sum_{\ell=1}^\infty \hat b_\ell \, x^{\ell-1} \ . 
\label{betaxsgd}
\eeq
The $n$-loop beta function $\beta_{x,n\ell}$ is defined by
Eq. (\ref{betaxsgd}) with the upper limit on the sum given by $\ell=n$. 
Explicitly, using the $\overline{\rm DR}$ scheme for $b_3$ and $b_4$, 
\beq
SYM: \quad \beta_{x,4\ell,\overline{\rm DR}} = -6x^2 (1+2x+7x^2+34x^3) \ . 
\eeq
It will be convenient to define a reduced ($rd$) beta function, $\beta_{x,rd}$:
\beq
\beta_{x,rd} \equiv -\frac{\beta_x}{2x^2 \, \hat b_1} = 
1 + \frac{1}{\hat b_1} \, \sum_{\ell=2}^\infty \hat b_\ell \, x^{\ell-1} \ . 
\label{betaxreduced}
\eeq
This separates off the factor that gives rise to a UV zero at $x=0$ so that we
can concentrate on the region of interest, namely the IR behavior.  The point
here is that both $\beta_{x,NSVZ}$ and $\beta_{x,\overline{\rm DR}}$ are
guaranteed to have the same UV behavior in the vicinity of the origin because
of the asymptotic freedom of the theory and the fact that the first two orders
in the loop expansion are scheme-independent.  The question is how well
they agree in the IR.  As with the full beta function, we also define the
$n$-loop truncation of $\beta_{x,rd}$, denoted $\beta_{x,rd,n\ell}$, as
Eq. (\ref{betaxreduced}) with the upper limit on the sum given by $\ell=n$;
this is thus a polynomial of degree $n-1$ in $x$.  In the $\overline{\rm DR}$
scheme, the four-loop reduced beta function is
\beq
SYM: \quad \beta_{x,rd,4\ell,\overline{\rm DR}} = 1 + 2x + 7x^2 + 34x^3 \ . 
\label{betaxreduced_4loop}
\eeq

We first analyze the zeros of $\beta_{x,rd,n\ell}$ for $2 \le n \le 4$.
Since each term is positive, it is clear that at the two-loop level and also,
in the $\overline{\rm DR}$ scheme, at the $n=3, \ 4$ loop level, the respective
$n$-loop reduced beta function has no physical zero. Specifically, 
the reduced two-loop beta function $\beta_{x,rd,2\ell}$ has only an 
unphysical zero away from the origin, at $x=-1/2$. With $b_3$ calculated in 
the $\overline{\rm DR}$ scheme, the reduced three-loop beta function
$\beta_{x,rd,3\ell}$ has an unphysical pair of complex-conjugate zeros, at
\beq
x=\frac{1}{7} \, (-1\pm \sqrt{6} \, i) = -0.14286 \pm 0.34993 i 
\label{sgd_xir_3loop}
\eeq
(Here and below, floating-point numbers are listed to the indicated accuracy.)
The reduced four-loop beta
function $\beta_{x,rd,4\ell,\overline{\rm DR}}$ has three unphysical roots, at
\beq
x=-0.3152, \quad x=0.05466 \pm 0.3005i \ .
\label{sgd_xir_4loop}
\eeq

We now compare the properties of $\beta_{x,\overline{\rm DR}}$ and
$\beta_{x,NSVZ}$ for the SYM theory.  The series expansions of these two beta
functions about $x=0$ are necessarily equal up to two-loop order inclusive,
since the beta function is scheme-independent up to and including this order.
Beyond two-loop order they differ, as is to be expected, since they are
calculated in different schemes.  An important question is whether, although
they differ in detail, these two beta functions at least exhibit qualitatively
similar physical properties.  To answer this question, we first express the
NSVZ beta function for the SYM theory in terms of $\beta_x$, obtaining
\beq
\beta_{x,NSVZ,SYM} = -\frac{6x^2}{1-2x} \ , 
\label{betax_nsvz_sym}
\eeq
so that the reduced NSVZ beta function for the SYM theory is 
\beq
\beta_{x,rd,NSVZ,SYM}=\frac{1}{1-2x} \ . 
\label{betax_reduced_nsvz_sym}
\eeq
This $\beta_{x,rd,NSVZ,SYM}=1/(1-2x)$ is in the exact form of a [0,1]
Pad\'e approximant, a property that will be used below. Clearly, 
$\beta_{x,rd,NSVZ,SYM}$ has no IR zero and, as is evident from Eq. 
(\ref{apole_nsvz}) (or equivalently, Eq. (\ref{betax_reduced_nsvz_sym})), it 
has a pole at $x=1/2$:
\beq
x_{pole,NSVZ}= \frac{1}{2} \ . 
\label{xpole_nsvz}
\eeq
Interestingly, the $\beta_{x,rd,n\ell,\overline{\rm DR}}$ functions at the 
$n=2, \ 3, \ 4$ loop levels all share the same property as $\beta_{x,NSVZ}$ 
in having no (physical) IR zero. This property will be discussed further below.

We next carry out our Pad\'e calculations and analysis.  In general, the
$n$-loop reduced beta function $\beta_{x,rd,n\ell}$ is a polynomial of degree
$n-1$ in $x$. At loop order $n \ge 3$, the coefficients in this function depend
on the scheme used for the calculation, and hence, where this is not obvious
from context, we shall indicate the scheme with an additional subscript. Since
all of the $[p,q]$ Pad\'e approximants that we calculate will apply to the beta
function in the $\overline{\rm DR}$ scheme, it is not necessary to indicate
this. We can thus calculate $[p,q]$ Pad\'e approximants of the form
\beq
[p,q]_{\beta_{x,rd,n\ell}} = 
\frac{1+\sum_{j=1}^p \, n_j x^j}{1+\sum_{k=1}^q d_k \, x^k}
\label{pqx}
\eeq
with
\beq
p+q=n-1 \ ,
\label{pqn}
\eeq
where the $n_j$ and $d_j$ are $x$-independent coefficients of the respective
polynomials in the numerator and denominator of
$[p,q]_{\beta_{x,rd,n\ell}}$.  (Our notation for Pad\'e approximants
follows the notation in, e.g., \cite{smpade}.)  Thus, as with
$\beta_{x,rd,n\ell}$ itself, each Pad\'e approximant is normalized so that
$[p,q]_{\beta_{x,rd,n\ell}}=1$ at $x=0$.  For a given
$\beta_{x,rd,n\ell}$, there are thus $n$ Pad\'e approximants, namely the
set
\beq
\{ \ [n-k,k-1]_{\beta_{x,rd,n\ell}} \ \} \quad {\rm with} \ 
1 \le k \le n \ .
\label{padeset}
\eeq
We shall generically denote one of the $p$ zeros of a 
$[p,q]]_{\beta_{x,rd,n\ell}}$ Pad\'e
approximant as $[p,q]_{zero}$ and one of the $q$ poles of this
approximant as $[p,q]_{pole}$; in each case, the value of $n$ is given by
Eq. (\ref{pqn}) as $n=p+q+1$ and will sometimes be omitted for brevity.

Since 
\beq
[n-1,0]_{\beta_{x,rd,n\ell}} = \beta_{x,rd,n\ell} \ , 
\label{polynomialpade}
\eeq
i.e., the $[n-1,0]_{\beta_{x,rd,n\ell}}$ Pad\'e approximant is identical to the
$n$-loop reduced beta function itself, whose zeros we have already analyzed, we
mainly restrict our consideration below to Pad\'e approximants
$[p,q]_{\beta_{x,rd,n\ell}}$ with $q \ne 0$.  Since $b_1$ and $b_2$ are
scheme-independent, it follows that for $n=1, \ 2$, the corresponding Pad\'e
approximants are scheme-independent. Note that for an arbitrary polynomial
$1+\sum_j f_j x^j$, the zero of the $[1,0]$ Pad\'e approximant, $1+f_1 x$,
denoted as $[1,0]_{zero}$, occurs at minus the value of the pole in the $[0,1]$
approximant $1/(1-f_1 x)$, denoted as $[0,1]_{pole}$, i.e.,
\beq
[1,0]_{zero} = -\frac{1}{f_1} = -[0,1]_{pole} \ . 
\label{polezerorel}
\eeq
We can also explore the correspondence between the pole at $x=1/2$ in
$\beta_{x,rd,NSVZ}$ and the structure of 
$\beta_{x,rd,n\ell,\overline{\rm DR}}$.  
Although $\beta_{x,rd,n\ell,\overline{\rm DR}}$ is a polynomial and
hence obviously has no poles, we can investigate whether $[p,q]$ Pad\'e
approximants to $\beta_{x,rd,n\ell,\overline{\rm DR}}$ with $q \ne 0$ share
properties in common with $\beta_{x,rd,NSVZ}$.

We shall address and answer the following
specific questions concerning the $[p,q]$ Pad\'e approximants:
\begin{enumerate}

\item Considering the $[p,q]$ Pad\'e approximants to 
$\beta_{x,rd,n\ell,\overline{\rm DR}}$, do the $[p,q]$ approximants with 
$p \ne 0$ exhibit a physical IR zero?

\item  Considering the $[p,q]$ Pad\'e approximants to 
$\beta_{x,rd,n\ell,\overline{\rm DR}}$, do the $[p,q]$ approximants with 
$q \ne 0$ exhibit a physical pole? 

\item If a given $[p,q]$ Pad\'e approximant with $q \ne 0$ does exhibit a
  physical pole, does this pole dominate the UV to IR evolution? This is the
  case if and only if this pole occurs closer to the origin $x=0$ than a
  physical IR zero.

\item If the answers to the previous two questions are affirmative, then is the
  value of the pole in the given $[p,q]$ Pad\'e approximant with $q \ne 0$
  close to the value $x_{IR,NSVZ}=1/2$? 

\item If the answers to questions 2 and 3 are affirmative, then, independent of
  whether the poles in the $[p,q]$ approximants with $q \ne 0$ are close to
  $x_{IR,NSVZ}=1/2$, do different approximants at least exhibit a stable
  physical pole?  That is, do these $[p,q]$ Pad\'e approximants with $q \ne 0$
  exhibit a stable physical pole?

\end{enumerate}
The answers to these questions elucidate how general and robust are the
properties of the SYM beta function calculated in different schemes, in
particular, the absence of an IR zero and the presence of a pole in the NSVZ
beta function.  We have already partially answered the first question, since we
have shown that there is no IR zero in the two-loop beta function and also none
in the three-loop or four-loop beta function in the $\overline{\rm DR}$ scheme,
in agreement with the absence of an IR zero in the NSVZ beta function for this
SYM theory.  We complete this first answer by examining $[p,q]$ Pad\'e 
approximants with both $p$ and $q$ nonzero and also address the questions
pertaining to a pole.

From the two-loop reduced beta function $\beta_{x,rd,2\ell}$, we can 
calculate one Pad\'e approximant with $q \ne 0$, namely 
\beq
[0,1]_{\beta_{x,rd,2\ell}} = \frac{1}{1-2x} \ . 
\label{betax_reduced_2loop_pade01_sym}
\eeq
This is the same as the reduced NSVZ beta function, 
$\beta_{x,rd,NSVZ,SYM}$ in Eq. (\ref{betax_reduced_nsvz_sym}), and hence
their poles are at the same location:
\beq
[0,1]_{pole,\beta_{x,rd,2\ell}} = \frac{1}{2} = x_{pole,NSVZ} \ . 
\label{sympade01_pole}
\eeq
Although $\beta_{x,rd,NSVZ,SYM}=1/(1-2x)$ was obtained by a sum
to infinite-loop order and hence is scheme-dependent, the pole in the [0,1]
Pad\'e approximant was derived from the two-loop beta function
$\beta_{x,rd,2\ell}$ and hence is scheme-independent. 

We proceed next to the comparison at the three-loop order. 
From the reduced three-loop reduced beta function, 
$\beta_{x,rd,3\ell,\overline{\rm DR}}$, we can 
calculate two Pad\'e approximants with $q \ne 0$, namely 
\beq
[1,1]_{\beta_{x,rd,3\ell}} = \frac{1-(3/2)x}{1-(7/2)x} 
\label{betax_reduced_3loop_pade11_sym}
\eeq
and
\beq
[0,2]_{\beta_{x,rd,3\ell}} = \frac{1}{(1+x)(1-3x)} \ . 
\label{betax_reduced_3loop_pade02_sym}
\eeq
As is evident from (\ref{betax_reduced_3loop_pade11_sym}), the 
$[1,1]_{\beta_{x,rd,3\ell}}$ Pad\'e has
a pole at $x=2/7=0.2857$ and a zero at $x=2/3$.  These are listed in Table
\ref{sympades}. As the
theory flows from the UV to the IR, $x$ increases from 0 and reaches the IR
pole at 2/7 before it reaches the zero, so the latter is not relevant to this
UV to IR evolution from weak coupling. The
$[0,2]_{\beta_{x,rd,3\ell}}$ Pad\'e exhibits an unphysical
pole at $x=-1$ and a physical pole at $x=1/3$. Since this Pad\'e has no zero,
the pole at $x=1/3$ again dominates the UV to IR evolution. 

From the four-loop reduced beta function 
$\beta_{x,rd,4\ell,\overline{\rm DR}}$, we can 
calculate three $[p,q]$ Pad\'e approximants with $q \ne 0$, namely 
\beq
[2,1]_{\beta_{x,rd,4\ell}} = \frac{1-(20/7)x-(19/7)x^2}{1-(34/7)x} \ , 
\label{betax_reduced_4loop_pade21_sym}
\eeq
\beq
[1,2]_{\beta_{x,rd,4\ell}}=\frac{1-(14/3)x}{1-(20/3)x+(19/3)x^2} \ , 
\label{betax_reduced_4loop_pade12_sym}
\eeq
and
\beq
[0,3]_{\beta_{x,rd,4\ell}} = \frac{1}{1-2x-3x^2-14x^3} \ . 
\label{betax_reduced_4loop_pade03_sym}
\eeq

The $[2,1]_{\beta_{x,rd,4\ell}}$ Pad\'e has zeros at 
$x=(1/19)(-10 \pm \sqrt{233} \ )$, i.e,
$x=0.2771$ and $x=-1.3297$, and a pole at $x=7/34=0.2059$.  
The $[1,2]_{\beta_{x,rd,4\ell}}$ Pad\'e has a zero at 
$x=3/14=0.2143$ 
and two poles, at $x=(1/19)(10 \pm \sqrt{43} \ )$, i.e., $x=0.8714$ and 
$x=0.1812$.  Finally, the 
$[0,3]_{\beta_{x,rd,4\ell}}$ Pad\'e has three poles, at 
$x=0.26481$ and $x=-0.23955 \pm 0.4608i$.  As with our other Pad\'e results for
the SYM theory, these are listed in Table \ref{sympades}.

\begin{table}
\caption{\footnotesize{Values of zeros and poles, in the variable $x$, 
of various Pad\'e approximants to $\beta_{x,rd,2\ell}$ and 
$\beta_{x,rd,n\ell,\overline{\rm DR}}$ with $n=3, \ 4$ for ${\cal N}=1$ 
supersymmetric Yang-Mills theory, SYM.  Results are given 
to the indicated floating-point accuracy.  The abbreviation NA means ``not
applicable''.}}
\begin{center}
\begin{tabular}{|c|c|c|c|c|} \hline\hline
$n$ & $[p,q]$ & zero(s)             &  pole(s)            \\ \hline
 2  & [1,0]   & $-1/2$              & NA                  \\
 2  & [0,1]   & NA                  & 1/2                 \\ \hline
 3  & [2,0]   & $0.143 \pm 0.350i$  & NA                  \\
 3  & [1,1]   & 2/3                 & $2/7=0.286$         \\
 3  & [0,2]   & NA                  & $-1$, \ 1/3         \\ \hline
 4  & [3,0]   & $-0.315$, \ $0.0547 \pm 0.3005i$ &  NA    \\
 4  & [2,1]   & $-1.330$, \ $0.277$ & $7/34 = 0.206$      \\
 4  & [1,2]   & $3/14=0.214$        & 0.181, \ 0.871      \\
 4  & [0,3]   & NA                  & 0.265, \ $-0.240 \pm 0.461i$    \\
\hline\hline
\end{tabular}
\end{center}
\label{sympades}
\end{table}

These results provide answers to the five questions that we posed above.
Concerning the first question, the [1,0] approximant to $\beta_{x,rd,2\ell}$
and the [2,0], and [3,0] approximants to $\beta_{x,rd,n\ell,\overline{\rm DR}}$
for $n=3, \ 4$ have no IR zero, in agreement with the NSVZ beta function.  
Although the [1,1] approximant to $\beta_{x,rd,3\ell\overline{\rm DR}}$ and
the [2,1] and [1,2] approximants to $\beta_{x,rd,4\ell,\overline{\rm DR}}$ do
have (physical) IR zeros, in each case, the IR zero occurs farther from the
origin than the pole in the respective approximant and hence does not directly
influence the UV to IR evolution.  Thus, the results from all of these Pad\'e
approximants to $\beta_{x,rd,n\ell,\overline{\rm DR}}$ for $2 \le n \le 4$ 
are in agreement with the NSVZ beta function for this SYM theory as regards the
absence of an IR zero affecting the UV to IR evolution.

The answer to the second question is yes; that is, all of the Pad\'e
approximants to the various $n$-loop reduced beta functions in the
$\overline{\rm DR}$ scheme, $\beta_{x,rd,n\ell,\overline{\rm DR}}$, up to $n=4$
loop order, agree with $\beta_{x,rd,NSVZ}$ as regards the property that they
exhibit a physical IR pole.  The answer to the third question is also yes; in
each case where a $[p,q]$ Pad\'e approximant to
$\beta_{x,rd,n\ell,\overline{\rm DR}}$ exhibits a physical pole, this pole
occurs closer to the origin than any physical zero(s) (if such a zero is
present at all) and hence dominates the UV to IR evolution. 

We come next to question 4, concerning the numerical agreement of the
(physical) pole in the $[p,q]$ Pad\'e approximants with $q \ne 0$ with the
position of the pole at $x=1/2$ in the NSVZ beta function.  To answer this
question, for each $[p,q]$ approximant to $\beta_{x,rd,n\ell,\overline{\rm
    DR}}$ with $q \ne 0$, one takes the pole among the $q$ poles at a physical
(positive real) value of $x$ (if there is such a pole) closest to the origin.
This is the IR pole for this approximant. As noted, this agreement is automatic
in the two-loop case, so the question really applies at the three-loop and
four-loop level. As is evident in Table \ref{sympades}, the two Pad\'e
approximants with $q \ne 0$ formed from the three-loop beta function, namely
[1,1] and [0,2], have poles at the respective values $x=0.286$ and
$x=0.333$. At the four-loop level, the values of the poles closest to the
origin in the [2,1], [1,2], and [0,3] Pad\'e approximants are $x=0.206,
\ 0.181, \ 0.265$, respectively.  None of these is particularly close to the
value $x=0.5$ of the pole in the NSVZ beta function.  This is in contrast with
the results that were obtained in \cite{bvh}-\cite{schl} (see also
\cite{ryttovschemes}) concerning the scheme dependence of the IR zero in the
beta function for a nonsupersymmetric asymptotically free non-Abelian gauge
theory calculated up to four-loop level; in these studies, it was found that
the position of this zero does not change very much as one applies various
scheme transformations to the results calculated \cite{bvh,ps} in the
$\overline{\rm MS}$ scheme.  Nevertheless, we can at least say that the values
of the (physical) pole in the various $[p,q]$ approximants to
$\beta_{x,rd,n\ell,\overline{DR}}$ with $n=3, \ 4$ do not differ from the value
$x=1/2$ in $\beta_{x,rd,NSVZ}$ by more than a factor of about 2.8.

Finally, we address the fifth question.  The importance of this question stems
from the fact that when one switches schemes, one does not expect a pole (or
zero) to occur at the same position as in another scheme, but at least
different $[p,q]$ Pad\'e approximants should yield a stable value of this pole,
especially as one calculates to progressively higher-loop order.  There are
thus two categories of comparisons that one can make here, namely comparing the
stability of the position of a pole appearing in $[p,q]$ Pad\'e approximants
for different loop orders $n$, and comparing this stability for a given
$n$-loop order in $[p,q]$ Pad\'e approximants with different $p$ and $q$
(satisfying $p+q=n-1$), at a high-enough order so that there are several
$[p,q]$ approximants with poles.  Regarding the comparison among
different loop orders, as is evident from Table \ref{sympades}, the values of
the poles range from the value from the two-loop result, which is automatically
equal to $x=0.5$, to a low of $x=0.181$ for the physical pole in the
$[1,2]_{\beta_{x,rd,4\ell}}$ Pad\'e, a factor of 2.8 smaller. 
Regarding the range of values of physical pole
positions from the Pad\'e approximants at a given loop order, the range is
given, at the three-loop order, by the ratio
\beq
\frac{[0,2]_{pole}}{[1,1]_{pole}} = \frac{7}{6} = 1.167 \ , 
\label{sym_pade02_pole_over_sym_pade11_pole}
\eeq
and, at the four-loop order, by two independent ratios,
which may be taken to be
\beq
\frac{[0,3]_{pole}}{[1,2]_{pole}} = 1.462 \ , 
\label{sym_pade03_pole_over_sym_pade12_pole}
\eeq
and
\beq
\frac{[1,2]_{pole}}{[2,1]_{pole}} = 0.88005 \ . 
\label{sym_pade12_pole_over_sym_pade21_pole}
\eeq
One also has 
\beq
\frac{[0,3]_{pole}}{[2,1]_{pole}} = 
\frac{[0,3]_{pole}}{[1,2]_{pole}} \, 
\frac{[1,2]_{pole}}{[2,1]_{pole}} = 1.286 \ . 
\label{sym_pade03_pole_over_sym_pade21_pole}
\eeq
Ideally, one would have hoped that this ratio, i.e.,
the scatter in the values of the IR pole positions, would decrease as the loop
order increased, but, at least up to four-loop order, it does not.

Summarizing the findings from our Pad\'e analysis for the SYM theory, the
results show excellent agreement between the beta function, calculated up to
four-loop order in the $\overline{\rm DR}$ scheme, and the NSVZ beta function,
concerning the absence of an IR zero that affects the UV to IR
evolution. Furthermore, the answers to questions 2 and 3 show that the Pad\'e
approximants to this beta function in the $\overline{\rm DR}$ scheme are
consistent with the existence of an IR pole that dominates the UV to IR
evolution, again in agreement with the NSVZ beta function.  The answer to the
fourth question can be interpreted as a consequence of the scheme-dependence of
a pole in a beta function.  The answer to the fifth question suggests that,
assuming that the beta function in the $\overline{\rm DR}$ scheme does, indeed,
encode evidence for a physical pole that dominates the UV to IR evolution in
this SYM theory, one must calculate this beta function to higher than four-loop
order in order for the $[p,q]$ Pad\'e approximants with $q \ne 0$ to yield a
stable value for the location of this zero.


\section{Supersymmetric SU($N_c$) Quantum Chromodynamics}
\label{sqcd}

In this section we investigate an asymptotically free vectorial gauge theory
with ${\cal N}=1$ supersymmetry, gauge group SU($N_c$),
and $N_f$ copies (flavors) of massless chiral superfields $\Phi_i$ and $\tilde
\Phi_i$, $i=1,...,N_f$, transforming according to the fundamental and conjugate
fundamental representations of SU($N_c$), with Young tableaux $\fund$ and
$\overline{\fund}$:
\beq
\Phi_i: \ \fund; \quad \tilde \Phi_i: \ \overline{\fund} \ , 
\quad {\rm with} \ i=1,...,N_f \ . 
\label{fundrep}
\eeq
This theory is often called supersymmetric quantum chromodynamics (SQCD), and
we shall also use this nomenclature, keeping in mind that the gauge group is
generalized from the actual SU(3) color group of real-world QCD to SU($N_c$).
We restrict our consideration to values $N_f \ne 0$ here, since if $N_f=0$, the
present theory reduces to a pure supersymmetric Yang-Mills gauge theory, which
we have already discussed above.  As we did with the SYM theory, we shall use
Pad\'e approximants to investigate the question of the extent to which the beta
function for this theory, as calculated in the $\overline{DR}$ scheme, exhibits
properties in agreement with the properties of the NSVZ beta function,
(\ref{beta_nsvz}).  


\subsection{Some General Properties}

We recall some basic well-known properties of this theory, many of which follow
as special cases of the general discussion in Sect. \ref{betafunction} 
for $R=\fund$.  For this case, the upper bound on $N_f$ imposed by the
condition of asymptotic freedom, Eq. (\ref{nfmax}), reads 
\beq
N_f < 3N_c \ . 
\label{nfmax_fund}
\eeq
The exact result (\ref{nfcr}) on the value of $N_f$ at the 
lower boundary of the IR non-Abelian Coulomb phase
reads \cite{nfintegral} 
\beq
N_{f,cr} = \frac{3N_c}{2} \ , 
\label{nfcr_fund}
\eeq
If $N_c$ is odd, this is only a formal result, since $N_{f,cr}$ must be
integral. Thus, the (chirally symmetric, deconfined) 
IR non-Abelian Coulomb phase is specified, from Eq. (\ref{nfnacp}), by
$N_f$ in the interval
\beq
\frac{3N_c}{2} < N_f < 3N_c \ . 
\label{nfnacp}
\eeq

For our present case with $R$ being the fundamental representation, 
Eq. (\ref{nfb2z}) specializes to 
\beq
N_{f,b2z} = \frac{3N_c}{2-N_c^{-2}} \ . 
\label{nfb2z_fund}
\eeq
Hence, the range of values of $N_f$ in Eq. (\ref{nfinterval}) where the 
two-loop beta function has an IR zero is \cite{nfintegral}
\beq
\frac{3N_c}{2-N_c^{-2}} < N_f < 3N_c \ . 
\label{nfinterval_fund}
\eeq
Numerical values of $N_{f,cr}$, $N_{f,b2z}$, $N_{f,b3z}$, and
$N_{f,b1z}=N_{f,max}$ were listed for $2 \le N_c \le 5$ in Table II of
Ref. \cite{bfs}. As was noted in \cite{bfs}, the value of $N_{f,b2z}$ in
Eq. (\ref{nfb2z_fund}) is greater (for all finite $N_c$) than the exactly known
lower boundary of the non-Abelian Coulomb phase in Eq. (\ref{nfcr_fund}).
Results for the values of the IR zero in the two-loop and three-loop beta
function, $\alpha_{IR,2\ell}$ and $\alpha_{IR,3\ell}$, were given in
\cite{bfs}.


\subsection{Calculations of Pad\'e Approximants}

We now proceed to calculate and analyze the Pad\'e approximants to the 
$n$-loop beta function for this SQCD theory. As before with the SYM theory, 
since our analysis concerns the behavior away from the UV fixed
point at $\alpha=0$, it is convenient to deal with the reduced beta function
defined by (\ref{betaxreduced}).  Because the beta function $\beta_{\alpha}$ is
known up to three-loop order, the reduced beta function has the form
\beq
\beta_{\alpha,rd,3\ell} = 1 + (b_2/b_1)a + (b_3/b_1)a^2 \ . 
\label{beta_reduced_3loop}
\eeq
Since $G={\rm SU}(N_c)$, it follows that $C_A=N_c$ and the 
variable $x$ in Eq. (\ref{xdef}) has the explicit form
\beq
x \equiv a \, N_c \equiv \frac{\xi}{4\pi} \ . 
\label{xdef_lnn}
\eeq

In general, the beta function and hence the Pad\'e approximants to it depend on
the two parameters $N_c$ and $N_f$.  It is natural to apply the Pad\'e analysis
to address the question of how the properties of the beta function calculated
in the $\overline{\rm DR}$ scheme compare with those of the NSVZ beta function
in the simplest context, namely the limit where the
(appropriately scaled) beta function depends only one one variable.
This is the 't Hooft-Veneziano or LNN (Large $N_c$ and $N_f$) limit 
\beqs
LNN: & & \ N_c \to \infty, \quad N_f \to \infty, \cr\cr
     & & {\rm with} \ r \equiv \frac{N_f}{N_c} \ {\rm fixed \ and \ finite} 
\label{lnn}
\eeqs
with $x(\mu) = a(\mu) \, N_c$ a finite function of the Euclidean scale $\mu$. 

Our constraint of asymptotic freedom implies $r < 3$.
We divide our analysis into two parts corresponding to two
subdivisions of this interval, namely the NACP interval
\beq
I_{r,NACP}: \quad \frac{3}{2} < r < 3  \ , 
\label{rinterval_nacp}
\eeq
where the UV to IR evolution leads to a non-Abelian Coulomb phase without any
spontaneous chiral symmetry breaking, and the remaining interval $0 < r < 3/2$.
In addition to simplifying the analysis of the beta function from dependence on
two variables to dependence on one variable, the LNN limit has the appeal that
the interval (\ref{rinterval_nacp}) in which the two-loop beta function has an
IR zero coincides with the interval leading to a non-Abelian Coulomb
phase. This is in contrast to the situation for general $N_c$ and $N_f$, in
which $b_2$ vanishes in the interior of the NACP.

In the LNN limit, one focuses on the scaled beta function, which is finite in
this limit.  For this we use the same notation, $\beta_x$, as in
Eq. (\ref{betax}), with it being understood that the LNN limit is taken, so
that
\beq
\beta_x = \lim_{LNN} \, \frac{dx}{dt} \ , 
\label{betax_lnn}
\eeq
equivalent to $\beta_\xi = \frac{d\xi}{dt}$.  The function $\beta_x$ has the
expansion (\ref{betaxsgd}) with 
\beq
\hat b_\ell \equiv \lim_{LNN} \, \frac{b_\ell}{N_c^\ell} \ . 
\label{bellhat_lnn}
\eeq
As before, we denote the $n$-loop truncation of Eq. (\ref{betax_lnn}) as
$\beta_{x,n\ell}$, and, where appropriate, we indicate the scheme used for loop
order $n \ge 3$ by a subscript, as $\beta_{x,n\ell,\overline{\rm DR}}$. 
From Eqs. (\ref{b1susy}) and (\ref{b2susy}), it follows that the 
scheme-independent scaled coefficients are 
\beq
\hat b_1 = 3-r 
\label{b1hat_lnn}
\eeq
and
\beq
\hat b_2 = 2(3-2r) \ . 
\label{b2hat_lnn}
\eeq
From the expression (\ref{b3susy}) for $b_3$ calculated in the 
$\overline{\rm DR}$ scheme, one has
\beq
\hat b_3 = 21-21r+4r^2 \ . 
\label{b3hat_lnn}
\eeq
Thus, 
\beq
\beta_{x,3\ell,\overline{\rm DR}} =
-2x^2 \Big [ (3-r) + 2(3-2r)x + (21-21r+4r^2)x^2 \ \Big ] \ , 
\label{betax_3loop}
\eeq
and hence
\beqs
& & \beta_{x,rd,3\ell,\overline{\rm DR}} = 1 + \frac{\hat b_2}{\hat b_1} \, x + 
\frac{\hat b_3}{\hat b_1} \, x^2 \cr\cr
& = & 1 + 2\Big ( \frac{3-2r}{3-r} \Big ) \, x + 
\Big ( \frac{21-21r+4r^2}{3-r} \Big ) \, x^2 \ . \cr\cr
& & 
\label{betax_reduced_3loop_lnn}
\eeqs
It will be convenient to define
\beq
D_s = -\hat b_3 = -21+21r-4r^2 \ . 
\label{ds}
\eeq
This polynomial $D_s$ has the property that 
\beq
D_s > 0 \quad {\rm for} \quad \frac{1}{8}(21 - \sqrt{105} \ ) < r < 
                              \frac{1}{8}(21 + \sqrt{105} \ ) \ , 
\label{dspos}
\eeq
i.e., for $1.3441 < r < 3.9059$, which 
includes all of the interval $I_{r,NACP}$. (If $r$ lies outside of the 
interval in Eq. (\ref{dspos}), then $D_s < 0$.)

We recall from \cite{bc} that if the two-loop beta function has an IR
zero, then, since this is a scheme-independent property, one may require a
physically acceptable scheme to maintain the existence of this IR zero at loop
level $n \ge 3$ and the condition that it should maintain it at the three-loop
level implies that $b_3 < 0$ (see the proof in Section II.E of \cite{bc}). This
condition is thus satisfied by the $\overline{\rm DR}$ scheme, since 
$\hat b_3 < 0$ for the interval $I_{r,NACP}$, where
$\beta_{x,2\ell}$ has an IR zero. 


\subsection{Analysis for Interval $r \in I_{r,NACP}$ }


\subsubsection{ IR Zero of $\beta_{x,rd,2\ell}$} 

At the two-loop level, $\beta_{x,rd,2\ell}$ has a (scheme-independent) IR 
zero at 
\beq
x_{IR,2\ell} = \frac{\xi_{IR,2\ell}}{4\pi} = \frac{(3-r)}{2(2r-3)} \ ,
\label{xir2loop_lnn}
\eeq
which is physical for $3/2 < r < 3$, i.e., for $r \in I_{r,NACP}$.  The value
of $x_{IR,2\ell}$ increases monotonically from 0 to arbitrarily large values as
$r$ decreases from 3 to 3/2 in the interval $I_{NACP}$.
Clearly, for the values of $r$ in the lower part of this interval, where
$x_{IR,2\ell}$ becomes large, the perturbative calculation that yielded the 
expression for $x_{IR,2\ell}$ cannot be reliably applied. 


\subsubsection{ IR Zero of $\beta_{x,rd,3\ell,\overline{\rm DR}}$} 

At the three-loop level, $\beta_{x,rd,3\ell,\overline{\rm DR}}$ 
has an IR zero at \cite{bfs,lnn}
\beq
x_{IR,3\ell} = \frac{-(2r-3) + \sqrt{C_s}}{D_s} \ , 
\label{xir_3loop}
\eeq
where
\beq
C_s = -54+72r-29r^2+4r^3 \ . 
\label{cs}
\eeq
The polynomial $C_s$ has only one real zero, at $r=1.3380$, 
and is positive (negative) for $r$ greater (less)
than this value. Thus, $C_s$ is positive for all $r \in I_{r,NACP}$.
Since  $\hat b_3 < 0$ for $r \in I_{r,NACP}$, it follows that 
\beq
x_{IR,3\ell} \le x_{IR,2\ell} \ , 
\label{xir_3loop_le_xir_2loop}
\eeq
as a special case of an inequality that was proved in \cite{bc} (see
Eq. (2.29) of \cite{bc}). The inequality (\ref{xir_3loop_le_xir_2loop}) is a
strict inequality except at the upper end of $I_{r,NACP}$ at $r=3$, where 
both $x_{IR,3\ell}$ and $x_{IR,2\ell}$ vanish. 


\subsubsection{Analysis of IR Zero Using Pad\'e Approximants} 

For $r \in I_{r,NACP}$, where the beta function has an IR zero, we address and
answer the following set of questions concerning the comparison of the 
three-loop beta function calculated in the $\overline{\rm DR}$ scheme, the
Pad\'e approximants to it, and the NSVZ beta function:

\begin{enumerate}

\item  Considering (i) the $n$-loop beta function and (ii) the $[p,q]$ Pad\'e 
 approximants to this beta function with $p \ne 0$, do these 
 exhibit a physical IR zero?

\item If (i) the $n$-loop beta function and (ii) the $[p,q]$ Pad\'e approximant
  to this beta function with $p \ne 0$ do exhibit a physical IR zero, does this
  IR zero dominate the UV to IR evolution? This is the case if and only if this
  IR zero occurs closer to the origin $x=0$ than a physical IR pole (if the
  latter is present in a $[p,q]$ Pad\'e with $q \ne 0$ ).

\item In each of the cases (i) and (ii), if the answers to the previous two
  questions are affirmative, then is the value of the IR zero close to the
  value $x_{IR,cfs,NSVZ} \equiv x_{IR,NSVZ}$ in $\beta_{x,rd,NSVZ}$, given in
  Eq.  (\ref{xir_nsvz})?

\item In each of the cases (i) and (ii), if the answers to questions 1 and 2
  are affirmative, then, independent of the closeness of the IR zero
  to $x_{IR,cfs,NSVZ}$, are the values at least close to each other? 

\item For the $[p,q]$ Pad\'e approximants with $q \ne 0$, if there is a
  physical pole, is its location near to the value $x=1/2$ in
  the NSVZ beta function? 

\end{enumerate}

We recall that the $[p,0]=[n-1]$ Pad\'e approximant is identical to the reduced
$n$-loop beta function $\beta_{x,rd,n\ell}$, as noted above in
Eq. (\ref{polynomialpade}).  As a special case of this, the two-loop reduced
beta function $\beta_{x,rd,2\ell}$ yields only one Pad\'e approximant with a
zero, namely [1,0], which coincides with $\beta_{x,rd,2\ell}$, itself, so no
further analysis is necessary.  The three-loop reduced beta function
$\beta_{x,rd,3\ell,\overline{\rm DR}}$ yields two Pad\'e approximants with 
$p \ne 0$, namely [2,0] and [1,1].  The [2,0] approximant coincides with
$\beta_{x,rd,3\ell,\overline{\rm DR}}$, which has already been analyzed.  We
calculate the [1,1] approximant to be
\beq
[1,1]_{\beta_{x,rd,3\ell}}=\frac{1-\Big [\frac{E_s}{2(3-r)(2r-3)} \Big ] \, x}
{1- \Big [ \frac{D_s}{2(2r-3)} \Big ] \, x } \ , 
\label{beta_x_reduced_3loop_lnn_pade11}
\eeq
where
\beq
E_s = -27+36r-17r^2+4r^3 \ . 
\label{es}
\eeq
The polynomial $E_s$ has only one real zero, at $r = 1.3118$ and is positive
(negative) for $r$ greater (less) than this value. Therefore, $E_s$ is positive
for all $r \in I_{r,NACP}$.  As $r$ decreases from 3 to 3/2, $E_s$ decreases
from 36 to 9/4. Thus, the $[1,1]_{\beta_{x,rd,3\ell,\overline{\rm DR}}}$ Pad\'e
approximant has a zero at
\beq
x_{[1,1],zero} = \frac{2(3-r)(2r-3)}{E_s} \ . 
\label{beta_x_reduced_3loop_lnn_pade11_zero}
\eeq
This is positive semidefinite for all $r \in I_{r,NACP}$; it 
vanishes at both ends of this interval and reaches a maximum at 
$r=1.8321$ (a zero of the function $81-198r+189r^2-72r^3+8r^4$), where it has
the value $x_{[1,1],zero}=0.23898$. In order for $x_{[1,1],zero}$ to be 
relevant for the UV
to IR evolution of the theory (from weak coupling in the UV), it is necessary
that if this Pad\'e approximant has a pole at a physical value of $x$, then
this pole must occur farther from the origin than the zero. Below, when we
analyze poles of the various Pad\'e approximants, we will show that this
condition is satisfied (although the distance between the zero and the pole
vanishes as $r \searrow 3/2$).  We thus denote 
\beq
x_{[1,1],zero} = x_{IR,3\ell,[1,1]} \ . 
\label{xir_3loop_pade11zero}
\eeq

We prove two inequalities.  First, 
\beq
x_{IR,3\ell,[1,1]} \le x_{IR,2\ell} \quad {\rm for} \ r \in I_{r,NACP} \ .
\label{xir_3loop_pade11_le_xir_2loop}
\eeq
This is proved by computing the difference
\beq
x_{IR,2\ell} - x_{IR,3\ell,[1,1]} = \frac{(3-r)^2 \, D_s}{2(2r-3) \, E_s} \ . 
\label{xir_2loop_minus_xir_3loop_pade11zero}
\eeq
This difference is positive semidefinite for $r \in I_{r,NACP}$, vanishing only
as $r \nearrow 3$ at the upper end of this interval. 

Second, we obtain the stronger inequality 
\beq
x_{IR,3\ell,[1,1]} \le x_{IR,3\ell} \quad {\rm for} \ r \in I_{r,NACP} \ .
\label{xir_3loop_pade11_le_xir_3loop}
\eeq
(This is a stronger inequality since $x_{IR,3\ell} \le x_{IR,2\ell}$, by 
(\ref{xir_3loop_le_xir_2loop}).) We have proved the inequality
(\ref{xir_3loop_pade11_le_xir_3loop}) by calculating the difference, 
$x_{IR,3\ell}-x_{IR,3\ell,[1,1]}$ and showing that it is positive semidefinite
for $r \in I_{r,NACP}$, vanishing only at $r=3$. Combining these inequalities,
we have
\beq
x_{IR,3\ell,[1,1]} \le x_{IR,3\ell} \le x_{IR,2\ell} \quad {\rm for} \ r 
\in I_{r,NACP} \ , 
\label{genineq_lnn} 
\eeq
with equality only at $r=3$, where all three terms in the inequality vanish. 


\subsubsection{IR Zero from NSVZ Beta Function}

Applying the LNN limit to the NSVZ beta function (\ref{beta_nsvz}) and
calculating the resultant $\beta_x$ in Eq. (\ref{betax_lnn}), we obtain
\beq
\beta_{x,NSVZ,LNN} = -2x^2 \, \bigg [ \frac{3-r(1+\gamma_m)}{1-2x} \bigg ] \ .
\label{betax_nsvz}
\eeq
Here,
\beq
\gamma_m = \sum_{\ell=1}^\infty \hat c_\ell x^\ell \ , 
\label{gammam_lnn}
\eeq
where the maximal scheme-independent coefficient in $\gamma_m$ is the one-loop
coefficient 
\beq
\hat c_1 = 2 \ . 
\label{c1hat}
\eeq
In terms of the closed-form ($cf$) and series ($s$) functions defined in
\cite{bfss}, this beta function can be expressed as
\beq
\beta_{x,NSVZ} = -2x^2 \hat b_1 \, f_{x,cf,NSVZ} \, f_{x,s,NSVZ} \ , 
\label{betax_nsvz_lnn}
\eeq
where
\beq
f_{x,cf,NSVZ} = \frac{1}{1-2x} 
\label{fcf_nsvz_lnn}
\eeq
and
\beq
f_{x,s,NSVZ} = 1- \frac{r \, \gamma_m}{\hat b_1} = 1-\frac{r \, \gamma_m}{3-r} \ ,
\label{fs_nsvz_lnn}
\eeq
where here the subscript $s$ connotes the dependence on the series 
(\ref{gammam_lnn}) for $\gamma_m$.

There are different approaches to calculating the IR zero of
$\beta_{x,NSVZ}$. If one expands it in a series in $x$ around $x=0$ and
calculates the resultant zero, one necessarily reproduces the one-loop and
two-loop results obtained starting from the original series expansion, since
these are scheme-independent.  This analysis was carried out in
\cite{bfs,bc,lnn}.  An alternate approach proposed and analyzed in \cite{bfss}
is to incorporate the information obtained from the summation to infinite-loop
order that yields the structure in
Eqs. (\ref{betax_nsvz_lnn})-(\ref{fs_nsvz_lnn}). Since the factor
$f_{x,cf,NSVZ}$ has no zero, one thus calculates the IR zero as the zero in
$f_{x,s,NSVZ}$.  Substituting the expansion of $\gamma_m$ to its maximal
scheme-independent order $\gamma_m=2x$, one thus solves the equation
$1-2rx/(3-r)=0$, obtaining
\beq
x_{IR,NSVZ} = \frac{\xi_{IR,NSVZ}}{4\pi} = \frac{3-r}{2r} \ , 
\label{xir_nsvz}
\eeq
As $r$ decreases from 3 to 3/2 in the interval $I_{r,NACP}$, this IR zero,
$x_{IR,NSVZ}$, increases from 0 to 1/2. The IR zero $x_{IR,NSVZ}$ has
much better behavior than $x_{IR,2\ell}$ in that it increases to a finite value
as $r$ decreases to the lower end of the interval $I_{r,NACP}$, while
$x_{IR,2\ell}$ diverges at this lower boundary of $I_{r,NACP}$. (As noted
above, this divergence is only formal, since the perturbative calculation that
yielded the expression for $x_{IR,2\ell}$ ceases to apply when the value of $x$
becomes too large.) 

In order for the IR zero $x_{IR,NSVZ}$ to be relevant
to the UV to IR evolution of the theory, it is necessary and sufficient that
this IR zero of the beta function should occur closer to the origin than the
pole in $\beta_{x,NSVZ}$, which occurs at $x=1/2$, as given in
Eq. (\ref{xpole_nsvz}). The requisite condition 
\beq
x_{IR,NSVZ} \le x_{pole,NSVZ} \quad {\rm for} \ \frac{3}{2} < r < 3 
\label{xir_le_xpole_nsvz}
\eeq
is satisfied, since, as we have observed above, $x_{IR,NSVZ} < 1/2$ in this
interval, $3/2 < r \le 3$.  As $r$ approaches the lower boundary of
$I_{r,NACP}$ at $r=3/2$, $x_{IR,NSVZ}$ approaches $x_{pole,NSVZ}$ from below.
The inequality (\ref{xir_le_xpole_nsvz}) is a strict inequality except at the
single point $r=3$, where both $x_{IR,NSVZ}$ and $x_{pole,NSVZ}$ vanish.

As $r$ decreases below 3/2 in the interval $0 < r < 3/2$, $x_{IR,NSVZ}$
increases monotonically above 1/2.  Thus, for $0 < r < 3/2$, this IR zero at
$x_{IR,NSVZ}$ occurs farther from the origin $x=0$ than the IR pole in
$\beta_{x,NSVZ}$ at $x=1/2$ and hence is not directly relevant to the UV to IR
evolution of the theory from weak coupling.

We next prove some additional inequalities.  First, 
\beq
x_{IR,NSVZ} \le x_{IR,2\ell} \quad {\rm for} \ r \in I_{r,NACP} \ . 
\label{xir_le_xir_2loop}
\eeq
This is proved by calculating the difference, which is 
\beq
x_{IR,2\ell}-x_{IR,NSVZ} = \frac{(3-r)^2}{2r(2r-3)} \ . 
\label{xir_2loop_minus_xir_nsvz}
\eeq
This is evidently positive-semidefinite, vanishing only at the upper end of the
interval $I_{r,NACP}$ at $r=3$, where both $x_{IR,2\ell}$ and 
$x_{IR,NSVZ}$ vanish.  Next, we obtain the stronger inequality, 
\beq
x_{IR,NSVZ} \le x_{IR,3\ell} \quad {\rm for} \ r \in I_{r,NACP} \ , 
\label{xir_le_xir_3loop}
\eeq
with equality only at $r=3$, where both $x_{IR,NSVZ}$ and 
$x_{IR,3\ell}$ both vanish. This is again proved by calculating the difference:
\beq
x_{IR,3\ell} - x_{IR,NSVZ} = \frac{63-78r+29r^2-4r^3+2r\sqrt{C_s}}
{2rD_s} \ .
\label{xir_3loop_minus_xir_nsvz}
\eeq
The denominator of this expression is positive for $r \in I_{r,NACP}$. In the
numerator, although the polynomial $63-78r+29r^2-4r^3$ is negative for 
$r \in I_{r,NACP}$, it is smaller in magnitude than the second term, 
$2r\sqrt{C_s}$, so the numerator is positive semidefinite in this interval, 
vanishing only at the upper end, at $r=3$. 

Comparing $x_{IR,NSVZ}$ with $x_{IR,3\ell,[1,1]}$, we find that
\beq
x_{IR,3\ell,[1,1]} - x_{IR,NSVZ} = \frac{(3-r)^2(r-1)(4r-9)}{2rE_s} \ . 
\label{xir_3loop_pade11zero_minus_xir_nsvz}
\eeq
Therefore, the relative size of $x_{IR,NSVZ}$ and $x_{IR,3\ell,[1,1]}$ is
reversed between upper and lower subsections of the interval $I_{r,NACP}$: 
\beq
x_{IR,NSVZ} \le x_{IR,3\ell,[1,1]} \quad {\rm if} \ 2.25 \le r \le 3 
\label{xir_nsvz_le_xir_3loop_upperinterval}
\eeq
(with equality only at $x=2.25$ and $r=3$), while
\beq
x_{IR,NSVZ} > x_{IR,3\ell,[1,1]} \quad {\rm if} \ 1.5 < r < 2.25 \ . 
\label{xir_le_xir_3loop_lowerinterval}
\eeq

We summarize these results in Table \ref{irzeros_lnn}.
The entries in Table \ref{irzeros_lnn} for 
$x_{IR,n\ell}$ with $n=2, \ 3$ are equivalent to the entries for
$\xi_{IR,n\ell}=4\pi x_{IR,n\ell}$ with $n=2, \ 3$ given in Table VII of
\cite{lnn}; the entries for $x_{IR,3\ell,[1,1]}$ are new here.  As is evident,
the numerical results in Table \ref{irzeros_lnn} obey the general inequalities
(\ref{xir_3loop_pade11_le_xir_2loop}) and (\ref{xir_3loop_pade11_le_xir_3loop})
that we have proved above, as well as the inequality
(\ref{xir_3loop_le_xir_2loop}) proved in \cite{bc}.
\begin{table}
\caption{\footnotesize{Values of the IR zero of the beta function in the LNN
    limit of the SQCD theory, as a function of $r \in I_{r,NACP}$. The entries
    in the columns are: (i) $x_{IR,2\ell}$, IR zero of the 2-loop beta function
    $\beta_{x,2\ell}$, (ii) $x_{IR,3\ell}$, IR zero of the 3-loop beta function
    $\beta_{x,3\ell,\overline{\rm DR}}$, (iii) the IR zero calculated from the
    Pad\'e approximant [1,1] to $\beta_{x,3\ell,\overline{\rm DR}}$, and (iv)
    $x_{IR,cfs,NSVZ} \equiv x_{IR,NSVZ}$ obtained from $\beta_{x,NSVZ}$. Since
    $x_{IR,2\ell}$ formally diverges as $r \searrow 1.5$, the perturbative
    calculation is not applicable (NA) there.}}
\begin{center}
\begin{tabular}{|c|c|c|c|c|c|} \hline\hline
$r$ & $x_{IR,2\ell}$ & $x_{IR,3\ell}$ & $x_{IR,3\ell,[1,1]}$ & 
$x_{IR,cfs,NSVZ}$  \\ \hline
 1.5  & NA      & 1.000  & 0.500   & 0.000   \\
 1.6  & 3.500   & 0.690  & 0.4375  & 0.162   \\
 1.7  & 1.625   & 0.529  & 0.382   & 0.220   \\
 1.8  & 1.000   & 0.424  & 0.238   & 0.333   \\
 1.9  & 0.6875  & 0.349  & 0.236   & 0.289   \\
 2.0  & 0.500   & 0.290  & 0.222   & 0.250   \\
 2.1  & 0.375   & 0.242  & 0.202   & 0.214   \\
 2.2  & 0.286   & 0.201  & 0.179   & 0.182   \\
 2.3  & 0.219   & 0.166  & 0.154   & 0.152   \\
 2.4  & 0.167   & 0.135  & 0.129   & 0.125   \\
 2.5  & 0.125   & 0.107  & 0.104   & 0.100   \\
 2.6  & 0.0909  & 0.0811 & 0.0801  & 0.0769  \\
 2.7  & 0.0625  & 0.0579 & 0.0576  & 0.05555 \\
 2.8  & 0.0385  & 0.0368 & 0.0367  & 0.0367  \\
 2.9  & 0.0179  & 0.0175 & 0.0175  & 0.0172  \\
 3.0  & 0       & 0      & 0       & 0       \\
\hline\hline
\end{tabular}
\end{center}
\label{irzeros_lnn}
\end{table}
%


\subsubsection{Poles of Pad\'e Approximants} 

Here we investigate the poles of the Pad\'e approximants to
$\beta_{x,rd,2\ell}$ and $\beta_{x,rd,3\ell,\overline{\rm DR}}$ in order to
answer the questions posed above.  At the two-loop level, from
$\beta_{x,rd,2\ell}$ we can obtain one $[p,q]$ Pad\'e approximant with $q \ne
0$, namely
\beq
[0,1]_{\beta_{x,rd,2\ell}}=\frac{1}
{1 + \Big [ \frac{2(2r-3)}{3-r}\Big ]x } \ . 
\label{beta_x_reduced_2loop_lnn_pade01}
\eeq
This has a pole at
\beq
x_{[0,1]_{pole}} = -\bigg [ \frac{3-r}{2(2r-3)} \bigg ] \ . 
\label{pade01_pole_lnn}
\eeq
Since we are considering $r \in I_{NACP}$, i.e., $3/2 \le r \le 3$, this occurs
at negative $x$ and hence is unphysical.  As a special case of the general
result (\ref{polezerorel}), we have $x_{[0,1]_{pole}} = -x_{IR,2\ell}$. So the
fact that $x_{IR,2\ell}$ is physical guarantees that this pole is irrelevant. 

We next proceed to the three-loop level. From 
$\beta_{x,rd,3\ell,\overline{\rm DR}}$ we can
obtain two $[p,q]$ Pad\'e approximants with $q \ne 0$. The first is the [1,1]
approximant, given in Eq. (\ref{beta_x_reduced_3loop_lnn_pade11}). This has a
pole at
\beq
x_{[1,1]_{pole}} = \frac{2(2r-3)}{D_s} \ . 
\label{beta_x_reduced_3loop_lnn_pade11_pole}
\eeq
We list values of $x_{[1,1]_{pole}}$ as a function of $r$ in Table
\ref{padepoles_lnn}.  As $r$ decreases from 3 to 3/2 in the interval
$I_{r,NACP}$, the position of this pole decreases monotonically from 1 to 0.
For a given $r \in I_{r,NACP}$, this pole occurs farther from the origin than
the zero, i.e. $x_{[1,1]_{pole}} \ge x_{[1,1]_{zero}}$. 
We show this by calculating the difference,
\beq
x_{[1,1]_{pole}}-x_{[1,1]_{zero}} = \frac{8(2r-3)^3}{D_s \, E_s} \ . 
\label{pade11pole_minus_zero}
\eeq
The right-hand side of (\ref{pade11pole_minus_zero}) is positive for $3/2 < r
\le 3$ in $I$ and vanishes as $r$ decreases to 3/2 at the lower end of this
interval $I$.  Thus, the $[1,1]_{\beta_{x,rd,3\ell}}$ approximant exhibits a
physical zero closer to the origin than the pole, and hence the pole is not
relevant to the UV to IR evolution described by this Pad\'e approximant. This
irrelevance of the pole is similar to what we found for the
$[0,1]_{\beta_{x,rd,2\ell}}$ Pad\'e approximant; indeed in that case, the pole
occurred at an unphysical, negative value of $x$.  The confluence of the pole
and the zero of the $[1,1]_{\beta_{x,rd,3\ell}}$ approximant as 
$r \searrow 3/2$ in $I$ reflects the fact that as $r \searrow 3/2$, 
$[1,1]_{x,rd,3\ell} \to 1$, independent of $x$.

\begin{table}
\caption{\footnotesize{Values of the IR pole(s) of the $[p,q]$ Pad\'e
    approximants, with $q=1$ and $q=2$, to $\beta_{x,2\ell}$ and 
    $\beta_{x,3\ell,\overline{\rm DR}}$, as functions of $r$, in the LNN 
    limit of the SQCD theory.  The entries in the columns are 
    (i) $x_{[0,1]_{pole}}$, pole of [0,1] approximant to
    $\beta_{x,rd,2\ell}$; (ii) $x_{[1,1]_{pole}}$, pole of [1,1] approximant to
    $\beta_{x,rd,3\ell,\overline{\rm DR}}$; (iii) $\{ x_{[0,2]_{pole}} \}$,
    poles of [0,2] approximant to $\beta_{x,rd,3\ell,\overline{\rm DR}}$.}}
\begin{center}
\begin{tabular}{|c|c|c|c|c|} \hline\hline
$r$ & $x_{[0,1]_{pole}}$ & $x_{[1,1]_{pole}}$ & $\{ x_{[0,2]_{pole}} \}$
  \\ \hline
 0.0  & 0.500      & 0.286    & $-1, \quad 0.333$ \\
 0.1  & 0.518      & 0.296    & $-1.034, \quad 0.345$ \\
 0.2  & 0.538      & 0.307    & $-1.070, \quad 0.358$ \\
 0.3  & 0.5625     & 0.319    & $-1.109, \quad 0.373$ \\
 0.4  & 0.591      & 0.332    & $-1.150, \quad 0.390$ \\
 0.5  & 0.625      & 0.348    & $-1.195, \quad 0.410$ \\
 0.6  & 0.667      & 0.366    & $-1.245, \quad 0.434$ \\
 0.7  & 0.719      & 0.387    & $-1.304, \quad 0.463$ \\
 0.8  & 0.786      & 0.414    & $-1.376, \quad 0.500$ \\
 0.9  & 0.875      & 0.449    & $-1.473, \quad 0.549$ \\
 1.0  & 1.000      & 0.500    & $-1.618, \quad 0.618$ \\
 1.1  & 1.1875     & 0.584    & $-1.876, \quad 0.727$ \\
 1.2  & 1.500      & 0.769    & $-2.519, \quad 0.940$ \\
 1.3  & 2.125      & 1.739    & $-11.368, \quad 1.790$ \\
 1.4  & 4.000      &$-0.7143$ & $0.303 \pm 1.527i$    \\
 1.5  & $\infty$   & 0        & $\pm i$               \\
 1.6  & $-3.500$   & 0.1695   & $-0.0808 \pm 0.748i$  \\
 1.7  & $-1.625$   & 0.255    & $-0.110 \pm 0.588i$   \\
 1.8  & $-1.000$   & 0.3125   & $-0.119 \pm 0.473i$   \\
 1.9  & $-0.6875$  & 0.359    & $-0.118 \pm 0.385i$   \\
 2.0  & $-0.500$   & 0.400    & $-0.111 \pm 0.314i$   \\
 2.1  & $-0.375$   & 0.440    & $-0.101 \pm 0.256i$   \\
 2.2  & $-0.286$   & 0.479    & $-0.0895\pm 0.208i$   \\
 2.3  & $-0.219$   & 0.521    & $-0.0770\pm 0.167i$   \\
 2.4  & $-0.167$   & 0.566    & $-0.0644\pm 0.132i$   \\
 2.5  & $-0.125$   & 0.615    & $-0.0519\pm 0.101i$   \\
 2.6  & $-0.0909$  & 0.671    & $-0.0400\pm 0.0753i$  \\
 2.7  & $-0.0625$  & 0.734    & $-0.0288\pm 0.0526i$  \\
 2.8  & $-0.0385$  & 0.807    & $-0.0184\pm 0.0328i$  \\
 2.9  & $-0.0179$  & 0.895    & $-0.00875\pm 0.0154i$  \\
 3.0  & 0          & 1        & 0                      \\
\hline\hline
\end{tabular}
\end{center}
\label{padepoles_lnn}
\end{table}

For the analysis of an IR zero, as was carried out in \cite{bfs,lnn,bfss}, the
[0,2] Pad\'e is not of interest, since it cannot reproduce an IR zero that
is present in the analysis of $\beta_{x,rd,3\ell,\overline{\rm DR}}$.  
However, it is of interest
for the questions that we address in this subsection. We calculate 
\beqs
& & [0,2]_{\beta_{x,rd,3\ell}} = \frac{1}
{1 + \Big [ \frac{2(2r-3)}{3-r} \Big ] \, x 
+ \Big [ \frac{E_s}{(3-r)^2} \Big ] \, x^2 } \ .
\cr\cr
& & 
\label{beta_x_reduced_3loop_lnn_pade02} 
\eeqs
Since the coefficients of both the $x$ and $x^2$ terms in the denominator of
$[0,2]_{\beta_{x,rd,3\ell}}$ are positive, this approximant clearly has no pole
for physical (non-negative) $x$.  Explicitly, the poles in
$[0,2]_{\beta_{x,rd,3\ell}}$ occur at
\beq
x_{[0,2],pole} = \frac{(3-r)\Big [ -(2r-3) \pm \sqrt{F_s} \ \Big ]}{E_s} \ , 
\label{beta_x_reduced_3loop_lnn_pade02poles}
\eeq
where 
\beq
F_s = 36-48r+21r^2-4r^3 \ . 
\label{fs}
\eeq
The polynomial $F_s$ has only one real zero, at $r = 1.3223$, and is positive
(negative) for $r$ less (greater) than this value. Hence, $F_s$ is negative for
all $r \in I_{r,NACP}$. We list values of
$x_{[0,2],pole}$ as a function of $r$ in Table \ref{padepoles_lnn}. 

These results answer the five questions that we posed above.  The answer to the
first question is yes, the quantities
$\beta_{x,rd,2\ell}=[1,0]_{\beta_{x,rd,2\ell}}$,
$\beta_{x,rd,3\ell}=[2,0]_{\beta_{x,rd,3\ell,\overline{\rm DR}}}$, and
$[1,1]_{\beta_{x,rd,3\ell,\overline{\rm DR}}}$ all exhibit (physical) IR zeros.
This property is in agreement with existence of an IR in the NSVZ beta 
function for $r \in I_{r,NACP}$. 
Second, in each case, the respective IR zero controls the UV to IR evolution, 
and this again agrees with the NSVZ beta function. Actually, the only case to 
check is the $[1,1]_{\beta_{x,rd,3\ell,\overline{\rm DR}}}$ Pad\'e approximant,
for which we have proved that the pole is always farther from the origin than
the IR zero. This is also evident from an inspection of the entries for zeros
and poles in Tables \ref{irzeros_lnn}.  and \ref{padepoles_lnn}.  

Concerning the third and fourth questions, for $r$ in the upper
part of the interval $I_{r,NACP}$, where the IR zero in the $n$-loop beta
function occurs at a rather small value, one expects that the values of this IR
zero calculated from this beta function itself, from the Pad\'e approximants to
the (reduced) beta function, and from the NSVZ beta function should agree, and
this expectation is borne out by the results, as listed in Table
\ref{irzeros_lnn}.  For example, for the illustrative value $r=2.5$,
$x_{IR,2\ell}=0.125$, $x_{IR,3\ell}=0.107$, $x_{IR,3\ell,[1,1]}=0.104$, and
$x_{IR,NSVZ}=0.100$.  Aside from the lowest-order, two-loop value, the last
three values of the IR zero are quite close to each other. As $r$
decreases in the interval $I_{r,NACP}$, the differences tend to grow
somewhat. Thus, for $r=2.0$, $x_{IR,2\ell}=0.500$, $x_{IR,3\ell}=0.290$,
$x_{IR,3\ell,[1,1]}=0.222$, and $x_{IR,NSVZ}=0.250$.  Again, aside from the
lowest-order, two-loop value, the last three values are within about 10 \% of
each other.

Finally, concerning the fifth question, pertaining to the pole in the $[p,q]$
Pad\'e approximants with $q \ne 0$, one should remark at the outset that since
this pole occurs farther from the origin than the IR zero, it does not directly
affect the evolution from weak coupling in the UV to the IR, so its precise
value is not directly relevant for this evolution. The values of the pole
position from the $[0,1]$, $[1,1]$ and $[0,2]$ Pad\'e approximants are listed
in Table \ref{padepoles_lnn}. As discussed in connection with
Eq. (\ref{polezerorel}), given the fact that the series expansion of the NSVZ
beta function must agree with the $\overline{\rm DR}$ to two-loop order and
given the identity (\ref{polezerorel}), it follows that since the two-loop beta
function (equivalently, the [1,0] Pad\'e) has a physical IR zero, the pole in
the [0,1] Pad\'e approximant to $\beta_{x,r,2\ell}$ must occur at a negative
and hence unphysical value of $x$.  At the three-loop level, the two relevant
two Pad'e approximants, [1,1] and [0,2] have poles at different values of $x$,
and only one, namely [1,1], has a pole at a physical value of $x$.
Furthermore, the position of this pole varies as a function of $r$, decreasing
from 1 at $r=3$ to 0 at $r=3/2$, in contrast to the pole in $\beta_{x,NSVZ}$,
which has a fixed value at $x_{pole,NSVZ}=1/2$.  More generally, even the
unphysical poles in the [0,1] and [0,2] Pad\'e approximants vary considerably
as functions of $r$.  Thus, these Pad\'e approximants do not exhibit evidence
of a stable pole.  Of course, these results do not preclude the possibility
that if the beta function in the $\overline{\rm DR}$ scheme could be calculated
to higher order, one might begin to see evidence of a stable pole in the
$[p,q]$ Pad\'e approximants with $q \ne 0$.


\subsection{Analysis for $0 < r < 3/2$}

In this interval of $r$, $\hat b_1$ and $\hat b_2$ have the same sign, so the
two-loop beta function does not have any IR zero, and this is the maximum
scheme-independent information that one has concerning the IR zero.  Hence,
using this two-loop beta function, one infers that as the scale $\mu$ decreases
from the deep UV to the IR, the gauge coupling continuously increases,
eventually exceeding the region where one can use perturbative methods to
calculate it reliably.  Formally, the three-loop beta function calculated in
the $\overline{\rm DR}$ scheme continues to exhibit an IR zero,
$x_{IR,3\ell,\overline{\rm DR}}$, given in Eq. (\ref{xir_3loop}), for a small
interval of $r$ below 3/2.  However, one cannot take this to be a physically
compelling result, in view of the fact that the maximal scheme-independent
information available (from the two-loop beta function) does not exhibit any IR
zero. Furthermore, as $r$ decreases from 3/2 to the zero of $D_s$ at
$r=(21-\sqrt{105} \ )/8 = 1.3441$ (see Eq. (\ref{dspos})),
$x_{IR,3\ell,\overline{\rm DR}}$ grows without bound, so that one can ignore
it, since the value of the IR zero is beyond the regime where one would
consider perturbation theory to be reliable.  Indeed, as is evident from the
expression for $[1,1]_{zero}$ in Eq. (\ref{xir_3loop_pade11zero}), the [1,1]
Pad\'e approximant to $\beta_{x,r,3\ell,\overline{\rm DR}}$ ceases to have a
physical IR zero as $r$ decreases through 3/2.  The absence of a (physical) IR
zero from a perturbatively reliable calculation using the beta function in the
$\overline{\rm DR}$ scheme for $0 < r < 3/2$ is in agreement with the
prediction from the NSVZ beta function; as discussed above, for $r$ in the
interval $0 < r < 3/2$, the formal zero at $x_{IR,NSVZ}$ lies farther from the
origin than the pole at $x=1/2$ and hence is not directly relevant to the UV to
IR evolution of the theory.

We proceed to address and answer the following questions for this interval
$0 \le r < 3/2$:

\begin{enumerate}

\item  Do the $[p,q]$ Pad\'e approximants with $q \ne 0$ exhibit a physical
  pole?

\item  If the answer to the first question is affirmative, then does this pole
occur closer to the origin than any IR zero (if such a zero is present) and
hence dominate the UV to IR evolution of the theory?

\item If the answers to the first two questions are affirmative, then is this
  IR pole close to the value $x_{pole,NSVZ}=1/2$ of the pole in the NSVZ 
beta function?

\item If the answers to the first two questions are affirmative, then, 
independent of whether the poles in different $[p,q]$ Pad\'e approximants 
with $q \ne 0$ are close to $x_{IR,NSVZ}=1/2$, do these different 
approximants at least exhibit a stable (physical) pole? 

\end{enumerate} 

In the present case, the relevant Pad\'e approximants are [0,1], [1,1], and
[0,2].  As is clear from the explicit expressions for these approximants and
their poles given above, the poles do not occur at a fixed value of $x$ and are
not, in general, equal to $x_{IR,NSVZ}$.  As is evident from 
(\ref{pade01_pole_lnn}), the pole in the [0,1] Pad\'e
approximant to $\beta_{x,r,2\ell}$ occurs at $x=1/2$ only if $r=0$ and
increases monotonically without bound as $r$ increases from 0 to 3/2. 
The pole in the [1,1] Pad\'e approximant to 
$\beta_{x,r,3\ell,\overline{\rm DR}}$ increases
monotonically as a function of $r$ from 0.286 at $r=0$ and diverges as $r$
approaches the value $(21-\sqrt{105} \ )/8 = 1.3441$ from below. In the small
interval $1.3441 < r < 3/2$, this pole occurs at negative $x$ and hence is
unphysical.  Only at the value $r=1$ is the position of this pole in the 
[1,1] Pad\'e equal to 1/2.  The [0,2] Pad\'e approximant to 
$\beta_{x,r,3\ell,\overline{\rm DR}}$ has two poles, one of which is always
unphysical. The other pole of the [0,2] Pad\'e occurs at an $r$-dependent value
that increases from 1/3 at $r=0$ and diverges as $r$ approaches the zero in
$E_s$ at $r=1.3118$ from below.  It is negative for $r$ in the small interval
$1.3118 < r < 1.3223$, where $r=1.3223$ is the point where $F_s$ has a zero;
finally, in the small interval $1.3223 < r < 3/2$, it is complex.  This pole in
the [0,2] Pad\'e is approximately equal to 1/2 for $r=0.8$.  

These results provide answers to the questions stated above.  The answer to the
first question is that over much of the interval $0 \le r < 3/2$, the [0,1],
[1,1], and [0,2] Pad\'e approximants do exhibit physical poles, but these all
vary as functions of $r$ and are not stable at any particular fixed value.
Second, since there is no robust, scheme-independent IR zero in the beta
function, these poles do dominate the UV to IR evolution. Third, the values of
the poles in the various Pad\'e approximants exhibit some scatter and are not,
in general, equal to (the $r$-independent) value $x_{IR,NSVZ}=1/2$. Concerning
the fourth question, although the pole in the [0,1] Pad\'e is not very close to
the pole in the [1,1] Pad\'e or the physical pole in the [0,2] Pad\'e, the
latter two poles are in fair agreement with each other.  For example (see Table
\ref{padepoles_lnn}) over much of the interval $0 \le r < 3/2$, the ratio of
the physical pole in the [0,2] Pad\'e divided by the pole in the [1,1] Pad\'e
is about 1.2, which is reasonably close to unity. One may interpret this as
indicating that, at least for the three-loop beta function calculated in the
$\overline{\rm DR}$ scheme for this theory, there is rough agreement, at about
the 20 \% level, between the pole in the [1,1] Pad\'e approximant and the
physical pole in the [0,2] Pad\'e approximant.


\section{Chiral Superfields in Symmetric and Antisymmetric Rank-2 Tensor 
Representations}
\label{t2}

\subsection{Beta Function and IR Zeros} 

Here we consider a (vectorial, asymptotically free) ${\cal N}=1$ supersymmetric
SU($N_c$) gauge theory with $N_f$ copies of massless chiral superfields
$\Phi_i$ and $\tilde \Phi_i$, $i=1,...,N_f$, transforming according to the
symmetric and antisymmetric rank-2 tensor representations and their conjugates.
We denote these symmetric and antisymmetric rank-2 tensor representations as
$S_2$ and $A_2$, respectively, with corresponding Young tableaux $\sym$ and
$\asym$.  While the $S_2$ theory is defined for all $N_c \ge 2$, the $A_2$
theory is defined for $N_c \ge 3$, since the $A_2$ representation is a singlet
if $N_c=2$.  We restrict our consideration to values $N_f \ne 0$ here, since if
$N_f=0$, the present theory reduces to a pure supersymmetric Yang-Mills gauge
theory, which we have already analyzed above. The one-loop and two-loop
coefficients in the beta function are (e.g., \cite{bfs})
\beq
b_1 = 3N_c -(N_c \pm 2)N_f \ ,
\label{b1t2}
\eeq
and
\beq
b_2 = 2 \Big [ 3N_c^2(1-N_f) \mp 8(N_c-N_c^{-1})N_f \Big ] \ .
\label{b2t2}
\eeq
where the upper and lower signs apply for the $S_2$ and $A_2$ theories,
respectively.  Evaluating Eq. (\ref{b3susy}), we obtain, for $b_3$ in the 
$\overline{\rm DR}$ scheme, 
\beqs
b_3 & = & 7N_c^3(N_f-1)(N_f-3) \pm 2N_c^2N_f(-33+17N_f) \cr\cr
    & + & 8N_cN_f(1+5N_f) \mp 24N_f(N_f-1) \cr\cr
    & - & 16N_c^{-1}N_f(2+3N_f)\pm 64N_c^{-2}N_f \ .
\label{b3t2}
\eeqs
It will often be convenient to refer to these two cases together as $T_2$
(standing for tensor, rank-2) with the above sign convention, and we shall do
so.  The one-loop coefficient decreases with increasing $N_f$ and passes
through zero with sign reversal for $N_f=N_{f,b1z,T_2}$, where
\beq
N_{f,b1z,T_2} = \frac{3N_c}{N_c \pm 2} \ . 
\label{nfb1z_t2}
\eeq
In the $S_2$ theory, $N_{f,b1z,S_2}$ increases monotonically from 3/2 for
$N_c=2$ , approaching the limiting value 3 from below as $N_c \to \infty$,
while in the $A_2$ theory, $N_{f,b1z,A_2}$ decreases monotonically from 9 for
$N_c=3$, approaching the limiting value 3 from above as $N_c \to \infty$.
The two-loop coefficient also decreases with increasing $N_f$ and passes
through zero with sign reversal for $N_f=N_{f,b2z,T_2}$, where
\beq
N_{f,b2z,T_2} = \frac{3N_c^2}{3N_c^2 \pm 8(N_c-N_c^{-1})} \ .
\label{nfb2z_t2}
\eeq

From the exact results recalled above, it follows that the lower boundary of 
the IR non-Abelian Coulomb phase is 
\beq
N_{f,cr,T_2} = \frac{N_{f,b1z,T_2}}{2} = \frac{3N_c}{2(N_c \pm 2)} \ .
\label{nfcr_t2}
\eeq
In the $S_2$ theory, $N_{f,cr}$ increases monotonically from 3/4 for
$N_c=2$, approaching 3/2 from below as $N_c \to \infty$, while in the 
$A_2$ theory, $N_{f,cr}$ decreases monotonically from 9/2 for $N_c=3$,
approaching 3/2 from above as $N_c \to \infty$.  Hence, the IR non-Abelian
Coulomb phase exists for (integral) $N_f$ values in the interval 
\beq
I_{NACP,(S_2,A_2)}: \quad 
\frac{3N_c}{2(N_c \pm 2)} < N_f < \frac{3N_c}{N_c \pm 2} \ .
\label{nf_nacp_interval_t2}
\eeq
In the $S_2$ theory, as noted in \cite{bfs}, $N_{f,b2z,S_2} < N_{f,cr,S_2}$, so
$b_2 < 0$ if $N_f$ is in the non-Abelian Coulomb interval
(\ref{nf_nacp_interval_t2}).  In the $A_2$ theory, 
$N_{f,b2z,A_2} < N_{f,cr,A_2}$ if $N_f=1$ or $N_f=2$, while 
$N_{f,b2z,A_2} > N_{f,cr,A_2}$ if $N_f=3$. 

Assuming that $N_f$ is in the respective ranges where $b_2 < 0$ in the 
$S_2$ and $A_2$ theories, the theory has an IR zero in the beta function at the
two-loop level, occuring at $a_{IR,2\ell,T_2}=-b_{1,T_2}/b_{2,T_2}$, i.e., 
\beq
a_{IR,2\ell,T_2} = \frac{3N_c-(N_c \pm 2)N_f}
{2[3N_c^2(N_f-1) \pm 8(N_c-N_c^{-1})N_f ]} \ .
\label{alfir2loop_t2}
\eeq
As noted above, this two-loop result is scheme-independent.  
To calculate the
IR zero of the NSVZ beta function using the corresponding maximal
scheme-independent information in $\gamma_m$, we use Eq. (\ref{c1}) 
for the one-loop term in $\gamma_m$ and thus solve the equation 
$b_{1,T_2}-2N_fT_f c_1a=0$, obtaining the result
\beq
a_{IR,NSVZ,T_2} = \frac{N_c[N_c(3-N_f) \mp 2N_f]}
{4N_f(N_c \pm 2)^2(N_c \mp 1)} \ . 
\label{air_nsvz_t2}
\eeq

To analyze the extent to which the perturbative beta functions of the $S_2$ and
$A_2$ theories, calculated in the $\overline{\rm DR}$ scheme, exhibit
similarities with the NSVZ beta function, we proceed to consider these
functions at the three-loop level. To carry out this analysis, we focus on the
case $N_c \to \infty$ and again work with the scaled beta function $\beta_x =
dx/dt$. In this limit, the beta functions $\beta_{x,S_2}$ and $\beta_{x,A_2}$
for the $S_2$ and $A_2$ theories become the same, and we shall denote the
resulting beta function as $\beta_{x,T_2}$.  Similarly, the intervals given in
Eq. (\ref{nf_nacp_interval_t2}) also become the same, reducing to
\beq
I_{NACP,T_2}: \quad \frac{3}{2} < N_f < 3 \quad {\rm for} \ N_c \to \infty
\label{interval_nacp_ncinf}
\eeq
Hence, the interval $I_{NACP,T_2}$ contains only one physical, integral value
of $N_f$ in this limit, namely $N_f=2$. 

With the definition (\ref{bellhat}), we have, for the (scheme-independent)
one-loop and two-loop rescaled coefficients $\hat b_\ell$, the results
\beq
\hat b_{1,T_2} = 3-N_f \ , 
\label{b1hat_t2}
\eeq
and
\beq
\hat b_{2,T_2} = -6(N_f-1) \ .
\label{b2hat_t2}
\eeq
Note that $\hat b_{2,T_2}$ vanishes for $N_f=1$, so that the two-loop beta
function has no IR zero for this value of $N_f$.

In general, the IR zero of the rescaled two-loop beta function $\beta_{x,T_2}$ is $x_{IR,T_2}=-\hat{b_{1,T_2}}/\hat b_{2,T_2}$, i.e., 
\beq 
x_{IR,2\ell,T_2} = \frac{3-N_f}{6(N_f-1)} \ . 
\label{xir_2loop_t2}
\eeq
The only value of $N_f$ for which this has a finite, nonzero value is $N_f=2$,
and for $N_f=2$, $x_{IR,2\ell,T_2}=1/6$.  In this limit, the rescaled NSVZ beta
function $\beta_{x,NSVZ,T_2}$ has an IR zero at
\beq
x_{IR,NSVZ,T_2} = \frac{3-N_f}{4N_f} \ .
\label{xir_nsvz_t2}
\eeq
For $N_f=2$, this has the value
\beq
x_{IR,NSVZ,T_2}=\frac{1}{8} \quad {\rm for} \ N_f=2 \ . 
\label{xir_nsvz_t2_nf2}
\eeq

In the $\overline{\rm DR}$ scheme, the three-loop coefficient in
$\beta_{x,T_2}$ is 
\beq
\hat b_{3,T_2} = -7(N_f-1)(3-N_f) \ . 
\label{b3hat_t2}
\eeq
Hence, if $N_f=1$, the three-loop beta function in the $\overline{\rm DR}$ 
scheme is the same as the (scheme-independent) two-loop beta function, in which
the two-loop coefficient also vanishes.  Thus, if $N_f=1$ here, the beta
function reduces simply to the one-loop term, which has no IR zero. 

The three-loop beta function, $\beta_{x,3\ell,T_2}$, for the $S_2$
and $A_2$ theories in this limit, with $\hat b_3$ calculated in the
$\overline{\rm DR}$ scheme, is
\beq
\beta_{x,3\ell,T_2} = -2x^2( \hat b_{1,T_2} + \hat b_{2,T_2} x 
+ \hat b_{3,T_2} x^2) \ . 
\label{betax_3loop_t2}
\eeq
From Eq. (\ref{betaxreduced}), it follows that the reduced 
three-loop beta function $\beta_{x,rd,3\ell,T_2}$ is
\beqs
\beta_{x,rd,3\ell,T_2} & = & 1 + \frac{\hat b_{2,T_2}}{\hat b_{1,T_2}}x + 
\frac{\hat b_{3,T_2}}{\hat b_{1,T_2}}x^2 \cr\cr
& = & 1 - \frac{6(N_f-1)}{3-N_f} \, x -7(N_f-1)x^2 \ . \cr\cr
& & 
\label{betaxred_3loop_t2}
\eeqs

If $N_f \ne 1$ (and $N_f \ne 3$), the equation $\beta_{x,rd,3\ell,T_2}=0$ has,
formally, two solutions, given by
\beqs
x_{IR,3\ell,T_2,\pm} & = & \frac{1}{7(3-N_f)} \, \bigg [ -3 \pm 
\sqrt{\frac{7N_f^2-33N_f+54}{N_f-1}} \ \bigg ] \ .
\cr\cr
& & 
\label{xir_3loop_t2}
\eeqs
For $N_f=2$, the solution with the $+$ sign in front of the square root is 
equal to $x_{IR,3\ell,T_2,+}=1/7=0.14286$, which is reasonably close to the
NSVZ result of 1/8 given in Eq. (\ref{xir_nsvz_t2_nf2}). (The solution with 
the minus sign in Eq. (\ref{xir_3loop_t2}) has the
unphysical negative value $-1$ and hence is not relevant.)  


\subsection{Pad\'e Approximants}

We next calculate the Pad\'e approximants to the two-loop and three-loop
reduced beta functions, $\beta_{x,rd,2\ell,T_2}$ and $\beta_{x,rd,3\ell,T_2}$,
respectively.  As before, these functions are identical to the [1,0] and [2,0]
approximants, respectively, so our analysis given above of the zeros of
$\beta_{x,rd,2\ell,T_2}$ and $\beta_{x,rd,3\ell,T_2}$ applies to these
approximants.  We list these results in Table \ref{t2pades} and proceed to
calculate and analyze the $[p,q]$ approximants with $q \ne 0$. With
$\beta_{x,rd,2\ell,T_2}$ we can only calculate one $[p,q]$ approximant with $q
\ne 0$, namely the [0,1] approximant.  We find
\beq
[0,1]_{\beta_x,rd,2\ell,T_2} = \frac{1}{1+\frac{6(N_f-1)}{3-N_f}x} \ . 
\label{betax_reduced_3loop_pade01_t2}
\eeq
Thus, as a special case of the relation (\ref{polezerorel}), 
this Pad\'e approximant has a pole at
\beqs
[0,1]_{T_2,pole} & = & -[1,0]_{T_2,zero} = -x_{IR,2\ell,T_2} \cr\cr
                 & = & -\frac{(3-N_f)}{6(N_f-1)} \ . 
\label{pade01_t2_pole}
\eeqs
For $N_f=2$, this pole occurs at $x=-1/6$. 

\begin{table}
\caption{\footnotesize{Values of zeros and poles, in the variable $x$, of 
    Pad\'e approximants to $\beta_{x,rd,2\ell}$ and
    $\beta_{x,rd,3\ell,\overline{\rm DR}}$ for the ${\cal N}=1$ supersymmetric
    gauge theory with $N_f=2$ copies of chiral superfields in the symmetric or
    antisymmetric rank-2 tensor representation and its conjugate, in the limit
    $N_c \to \infty$. Results are given to the indicated floating-point
    accuracy.  The abbreviation NA means ``not applicable''.}}
\begin{center}
\begin{tabular}{|c|c|c|c|c|} \hline\hline
$n$ & $[p,q]$ & zero(s)             &  pole(s)               \\ \hline
 2  & [1,0]   & $1/6=0.167$         & NA                     \\
 2  & [0,1]   & NA                  & $-1/6=-0.167$          \\ \hline
 3  & [2,0] & $-1$, \ $1/7=0.143$   & NA                     \\
 3  & [1,1]   & $6/43=0.1395$       & $6/7=0.857$            \\
 3  & [0,2]   & NA                  & $-0.0698 \pm 0.1356i$  \\ 
\hline\hline
\end{tabular}
\end{center}
\label{t2pades}
\end{table}

At the three-loop level, assuming that $N_f \ne 1$ (where the three-loop and
two-loop coefficients vanish), we calculate, for the 
[1,1] approximant, the result
\beq
[1,1]_{\beta_x,rd,3\ell,T_2} = 
\frac{1-\frac{(7N_f^2-6N_f+27)}{6(3-N_f)}x} 
{1-\frac{7(3-N_f)}{6}x} \ . 
\label{betax_reduced_3loop_pade11_t2}
\eeq
The fact that $\hat b_2$ and $\hat b_3$ vanish at $N_f=1$ is reflected in the
property that this [1,1] Pad\'e approximant reduces to unity at this value of
$N_f$.  For $N_f \ne 1$, this approximant has a IR zero at
\beq
[1,1]_{T_2,zero} = \frac{6(3-N_f)}{7N_f^2-6N_f+27}  \ . 
\label{pade11_t2_zero}
\eeq
and a pole at
\beq
[1,1]_{T_2,pole} = \frac{6}{7(3-N_f)} \ . 
\label{pade11_t2_pole}
\eeq
The polynomial $7N_f^2-6N_f+27$ is positive for all (real) $N_f$.  We note the
inequality 
\beq
[1,1]_{T_2,zero} < [1,1]_{T_2,pole} \ . 
\label{pade11t2_inequality}
\eeq
This is proved by calculating the difference, 
\beqs
& & 
[1,1]_{T_2,pole}-[1,1]_{T_2,zero} = \frac{6^3(3-N_f)}{7(N_f-1)(7N_f^2-6N_f+27)}
\ . \cr\cr
& & 
\label{pade11t2_polezero_diff}
\eeqs
Evidently, since $N_f \ne 0, \ 1$ and $N_f < 3$, the right-hand side of 
(\ref{pade11t2_polezero_diff}) is positive-definite. 
For $N_f=2$, we have $[1,1]_{T_2,zero}=6/43=0.1395$ and 
$[1,1]_{T_2,pole}=6/7=0.8571$. These values are listed in Table \ref{t2pades}.

We compute the [0,2] Pad\'e approximant to $\beta_{x,rd,3\ell,T_2}$ to be
\beqs
& & [0,2]_{\beta_x,rd,3\ell,T_2} = \frac{1}
{1 + \frac{6(N_f-1)}{(3-N_f)}x + \frac{(N_f-1)(7N_f^2-6N_f+27)}{(3-N_f)^2}x^2}
\ . \cr\cr
& & 
\label{betax_reduced_3loop_pade02_t2}
\eeqs
This has, formally, two poles, at the values 
\beqs
[0,2]_{T_2,pole} & = & \frac{(3-N_f)}{(7N_f^2-6N_f+27)} \, \bigg [ -3 \cr\cr
& \pm & \sqrt{\frac{-7N_f^2+15N_f-36}{N_f-1}} \ \ \bigg ] \ . 
\label{pade02_t2_poles}
\eeqs
However, recalling that $N_f \ne 0, \ 1$, one sees that these poles are both
unphysical, because the polynomial in the square root, $-7N_f^2+15N_f-36$, is
negative-definite.  Explicitly, for $N_f=2$, Eq. (\ref{pade02_t2_poles}) yields
the pole values $x=(1/43)(-3 \pm \sqrt{34} \, i \ ) = -0.0698 \pm 0.1356i$, as
is listed in Table \ref{t2pades}. 

Let us summarize our results for these theories with a nonzero number, $N_f$,
of copies of chiral superfields in the $S_2$ or $A_2$ representation and their
respective conjugates, in the limit $N_c \to \infty$.  In this limit, $N_f < 3$
for asymptotic freedom.  For $N_f=1$, the (scheme-independent) two-loop term in
the beta function $\beta_{x,T_2}$ vanishes, as does the three-loop term with
the latter calculated in the $\overline{\rm DR}$ scheme, so that the beta
function, calculated to these loop orders, does not contain any IR zero or any
indication, via Pad\'e approximants, of a pole.

For this theory with $N_f=2$, our analysis of the two-loop and three-loop
$\overline{\rm DR}$ beta functions, and the [1,1] Pad\'e approximant to the
latter all give evidence of an IR zero, with respective values 1/6, 1/7, and
6/43=0.1395, which decrease monotonically, approaching the the value
$x_{IR,NSVZ,T_2}=1/8$ from the NSVZ beta function.  Thus, we find reasonably
good agreement between these IR zeros calculated in different schemes. We come
next to the questions of whether, for this theory, the $[p,q]$ Pad\'e
approximants with $q \ne 0$ to the two-loop and three-loop beta function give
some indication of a (physical) pole and whether this pole occurs at a value
close to the value $x=1/2$ in the NSVZ beta function.  Our result is that
neither the [0,1] nor the [0,2] Pad\'e approximants has any physical pole,
while the [1,1] Pad\'e does exhibit a pole, although it is roughly twice the
value $x=1/2$.  As in the case of the theory containing chiral superfields in
the $\fund$ and $\overline{\fund}$ representations, this suggests that it is
necessary to calculate the beta function in the $\overline{\rm DR}$ scheme to
higher-loop order and calculate higher-order Pad\'e approximants to test for
indications of a pole.


\section{Conclusions}
\label{conc}

In this paper we have studied asymptotically free vectorial SU($N_c$) gauge
theories with ${\cal N}=1$ supersymmetry, including both pure gluonic
supersymmetric Yang-Mills theory and theories with $N_f$ copies of a pair of
chiral superfields in respective representations $R$ and $\bar R$, where $R$ is
the fundamental representation and the symmetric and antisymmetric rank-2
tensor representation of SU($N_c$).  We have calculated Pad\'e approximants to
the beta functions for these theories in the $\overline{\rm DR}$ scheme up to
four-loop order for the SYM theory and up to three-loop order for the theories
with matter superfields and have compared results for IR zeros and poles with
results from the NSVZ beta function.  For the pure supersymmetric YM
theory, the Pad\'e results for the four-loop beta function calculated in the
$\overline{\rm DR}$ scheme show strong evidence for a pole, in good agreement
with the NSVZ beta function.

For the theory with chiral superfields in the fundamental representation, in
the large-$N_c$, large-$N_f$ limit with $3/2 < r < 3$, where $r=N_f/N_c$, the
theory evolves from weak coupling in the UV to a non-Abelian Coulomb phase in
the IR and exhibits an IR fixed point in the renormalization group. Our
analysis of the two-loop, three-loop beta function and the [1,1] Pad\'e
approximant to the latter yields robust evidence for the IR zero, with values
in reasonable agreement with the value obtained from the NSVZ beta function,
taking into account the difference between the $\overline{\rm DR}$ and NSVZ
schemes.  With regard to the question whether the Pad\'e approximants of the
beta function in the $\overline{\rm DR}$ scheme indicate a pole, as is present
in the NSVZ beta function, we find that of the three Pad\'e approximants
analyzed here ([0,1], [1,1] and [0,2]), only one, namely the [1,1] approximant,
yields a physical pole, and the $x$ value of this pole varies considerably as a
function of $r$, in contrast to the pole in the NSVZ beta function, which is a
constant at $x=1/2$, independent of $r$. With $r$ in the lower interval $0 < r
< 3/2$, the [0,1] and [1,1] approximants yield physical poles, and one of the
two poles in the [0,2] approximant is physical, but again, the $x$ values of
these poles vary considerably as functions of $r$, in contrast to the fixed
pole in the NSVZ beta function.

In the theory with chiral superfields in rank-2 tensor representations, in the 
$N_c \to \infty$ limit, if $N_f=1$, both the two-loop and the 
$\overline{\rm DR}$ three-loop coefficients vanish, while if
$N_f=2$, the various Pad\'e approximants yield an IR zero in
the beta function in agreement with the NSVZ beta function, but only one, 
namely the [1,1] approximant, yields a physical pole. 

Our calculations provide a quantitative measure, for these various
supersymmetric theories, of how well finite-order perturbative results
calculated in the $\overline{\rm DR}$ scheme reproduce the properties of the
NSVZ beta function.  Our results indicate qualitative and rough quantitative
agreement between the $\overline{\rm DR}$ and NSVZ beta functions for the
theories and matter superfield contents where these exhibit an IR zero.  Our
analysis of Pad\'e approximants to the $\overline{\rm DR}$ beta function for
the pure gluonic Yang-Mills theory show qualitative consistency with the IR
pole that is present in the NSVZ beta function.  For the supersymmetric
theories with matter superfields, our results suggest that it may be 
necessary to calculate the beta function to higher-loop order in the
$\overline{\rm DR}$ scheme in order to test consistency with the pole in the
NSVZ beta function.

\begin{acknowledgments}
This research was partially supported by the grant NSF-PHY-13-16617. 
\end{acknowledgments}



\begin{thebibliography}{99}

\bibitem{nsvz}
%
V. A. Novikov, M. A. Shifman, A. I. Vainshtein, and V. I. Zakharov,
Nucl. Phys. B {\bf 229}, 381, 407 (1983); Phys. Lett. B {\bf 166}, 329 (1986);
M. A. Shifman and A. I. Vainshtein, Nucl. Phys. B {\bf 277}, 456 (1986).

\bibitem{sv}
M. A. Shifman, Prog. Part. Nucl. Phys. {\bf 39}, 1 (1997). 

\bibitem{ks}
I. I. Kogan and M. Shifman, Phys. Rev. Lett. {\bf 75}, 2085 (1995). 

\bibitem{seiberg}
N. Seiberg, Phys. Rev. D {\bf 49}, 6857 (1994); 
K. A. Intriligator and N. Seiberg, Nucl. Phys. B {\bf 431}, 551 (1994);
N. Seiberg, Nucl. Phys. B {\bf 435}, 129 (1995).

\bibitem{susyreview}
A review is M. Shifman, Prog. Part. Nucl. Phys. {\bf 39}, 1 (1997) 
[hep-th/9704114]. 

\bibitem{b12susy}
S. Ferrara and B. Zumino, Nucl. Phys. B {\bf 79}, 413 (1974); 
D. R. T. Jones, Nucl. Phys. B {\bf 87}, 127 (1975). 

\bibitem{b3susy}
%
M. Machacek and M. Vaughn, Nucl. Phys. B {\bf 222}, 83 (1983);
A. J. Parkes and P. C. West, Phys. Lett. B {\bf 138}, 99 (1984);
Nucl. Phys. B {\bf 256}, 340 (1985);
D. R. T. Jones and L. Mezincescu, Phys. Lett. B {\bf 136}, 242 (1984); 
Phys. Lett. B {\bf 138}, 293 (1984);
I. Jack, D. R. T. Jones, and C. G. North, Nucl. Phys. B {\bf 473}, 308 (1996). 

\bibitem{b4sym}
I, Jack, D. R. T. Jones, and A. Pickering, Phys. Lett. B {\bf 435}, 61 (1998).

\bibitem{drbar} 
W. Siegel, Phys. Lett. B {\bf 84}, 193 (1979); 
W. Siegel, Phys. Lett. B {\bf 94}, 37 (1980); 
D. M. Capper, D. R. T. Jones, and P. van Nieuwenhuizen, Nucl. Phys. B 
{\bf 167}, 479 (1980); W. St\"ockinger, JHEP 0503, 076 (2005).

\bibitem{bfs}
T. A. Ryttov, R. Shrock, Phys. Rev. D {\bf 85}, 076009 (2012).

\bibitem{bc}
R. Shrock, Phys. Rev. D {\bf 87}, 105005 (2013).

\bibitem{lnn}
R. Shrock, Phys. Rev. D {\bf 87}, 116007 (2013).

\bibitem{jackjones}
I. Jack, D. R. T. Jones, and C. G. North, Nucl. Phys. B {\bf 486}, 479 (1997);
I. Jack, D. R. T. Jones, and A. Pickering, Phys. Lett. B {\bf 435}, 61 (1998).

\bibitem{sd}
T. Appelquist, A. Nyffeler, and S. B. Selipsky, Phys. Lett. B {\bf 425}, 300
(1998).

\bibitem{cems}
F. A. Chishtie, V. Elias, V. A. Miransky, and T. G. Steele, Prog. Theor. Phys.
{\bf 104}, 603 (2000).

\bibitem{elias}
V. Elias, J. Phys. G {\bf 27}, 217 (2001). 

\bibitem{ryttov07}
T. A. Ryttov and F. Sannino, Phys. Rev. D {\bf 76}, 105004 (2007).

\bibitem{ss}
F. Sannino and J. Schechter, Phys. Rev. D {\bf 82}, 096008 (2010);
A. L. Kataev and K. V. Stepanyantz, Theor. Math. Phys. {\bf 181}, 1531 (2014).

\bibitem{casimir}
%
The Casimir invariants $C_R$ and $T_R$ are defined in the standard way as
$\sum_a \sum_j {\cal D}_R(T_a)_{ij} {\cal D}_R(T_a)_{jk} = C_R \delta_{ik}$ and
$\sum_{i,j} {\cal D}_R(T_a)_{ij} {\cal D}_R(T_b)_{ji} = T_R \delta_{ab}$, where
$R$ is the representation with matrix ({\it Darstellung}) ${\cal D}_R(T_a)$,
and $T_a$ are the generators of $G$, so that for SU($N_c$), $C_A=N_c$ for the
adjoint ($A$) and $T_{fund}=1/2$ for the fundamental representation, etc. $C_f$
denotes $C_R$ for the fermion representation.

\bibitem{nfintegral}
%
This and other expressions for special values of $N_f$ involve an implicit
generalization of $N_f$ from nonzero integers to nonzero real numbers.  It is
understood implicitly that if a formal expression for a value of $N_f$
evaluates to a non-integral (real) value, one infers an appropriate integral
value of $N_f$ from it, namely the closest integer larger or smaller than the
expression, depending on the context.

\bibitem{smpade}
I-H. Lee and R. E. Shrock, Phys. Rev. B {\bf 36}, 3712 (1987); 
I-H. Lee and R. E. Shrock, J. Phys. A {\bf 21}, 3139 (1988); 
V. Matveev and R. Shrock, J. Phys. A {\bf 28}, 1557 (1995). 

\bibitem{bfss}
R. Shrock, Phys. Rev. D {\bf 91}, 125039 (2015). 

\bibitem{bvh} 
T. A. Ryttov, R. Shrock, Phys. Rev. D {\bf 83}, 056011 (2011).

\bibitem{ps}
C. Pica, F. Sannino, Phys. Rev. D {\bf 83}, 035013 (2011).

\bibitem{scc}
T. A. Ryttov and R. Shrock, Phys. Rev. D {\bf 86}, 065032 (2012);
T. A. Ryttov and R. Shrock, Phys. Rev. D {\bf 86}, 085005 (2012).

\bibitem{sch2}
R. Shrock, Phys. Rev. D {\bf 88}, 036003 (2013); 
R. Shrock, Phys. Rev. D {\bf 90}, 045011 (2014).

\bibitem{schl}
G. Choi and R. Shrock, Phys. Rev. D {\bf 90}, 125029 (2014). 

\bibitem{ryttovschemes}
T. A. Ryttov, Phys. Rev. D {\bf 89}, 056001 (2014);
T. A. Ryttov, Phys. Rev. D {\bf 90}, 056007 (2014).

\end{thebibliography}
\end{document}